\documentclass[twocolumn,groupedaddress,
superscriptaddress,preprintnumbers,
amsmath,amssymb,
aps,prb
]{revtex4-2}
\usepackage{graphicx}
\usepackage{dcolumn}
\usepackage{bm}%
\usepackage[colorlinks=true,citecolor=blue]{hyperref}
\DeclareUnicodeCharacter{2032}{\ensuremath{'}}
\hypersetup{colorlinks=true,citecolor=blue,linkcolor=red,urlcolor=blue}

\usepackage{float}
\usepackage{amsmath}

\usepackage{bm}
\newcommand{\beq}{\begin{equation}}
\newcommand{\eeq}{\end{equation}}
\newcommand{\beqnar}{\begin{eqnarray}}
\newcommand{\eeqnar}{\end{eqnarray}}
\newcommand{\bfig}{\begin{figure}}
\newcommand{\efig}{\end{figure}}

\usepackage{color}
\usepackage[dvipsnames]{xcolor} 
\usepackage{changes}

\begin{document}
\title{Time-dependent Berry curvature and quantum metric of Floquet-Bloch states}

\author{S. Sajad Dabiri}
\affiliation{Department of Physics, Zhejiang Normal University, Jinhua 321004, China}
\author{Reza Asgari}
\affiliation{Department of Physics, Zhejiang Normal University, Jinhua 321004, China}
\affiliation{School of Quantum Physics and Matter, Institute for Research in Fundamental Sciences (IPM), Tehran 19395-5531, Iran}
\date{\today}
\vspace{1cm}
\newbox\absbox
\begin{abstract}
The quantum geometry of Bloch bands, characterized by the Berry curvature and the quantum metric, underpins a wide range of linear and nonlinear responses in static systems. Here, we extend this framework to periodically driven (Floquet) systems by introducing a time-dependent Berry curvature and quantum metric defined directly in the Floquet-Bloch basis. We derive optical sum rules that relate the Fourier components of these geometric quantities to the optical conductivity and demonstrate that, under ideal Floquet-band occupations, the first-order DC Hall and longitudinal responses at harmonic frequencies vanish identically. We further introduce a mixed Berry curvature involving time and momentum derivatives, which gives rise to a non-adiabatic quantized charge-pumping mechanism that occurs naturally during each driving period without requiring adiabatic evolution. In addition, we identify the time-domain quantum metric as a measure of the energy fluctuations of a Floquet band and interpret its mixed components as quantifying polarization-energy correlations. A comprehensive symmetry analysis reveals how time-reversal, sublattice (chiral), particle-hole, inversion, rotational, and reflection symmetries constrain the time-dependent quantum geometric tensor and its associated topological invariant. Numerical simulations of the Rudner-Lindner-Berg-Levin model and a fully symmetric Floquet model confirm the analytical predictions. These results establish the time-dependent quantum geometric tensor as a unified framework for describing the geometric, topological, and dynamical properties of periodically driven quantum systems, with direct implications for optical spectroscopy, quantum transport, and topological charge-pumping experiments.
\end{abstract}
\maketitle

\section{Introduction}

The interplay between time-periodic driving and topological band theory has given rise to the rapidly growing field of Floquet topological matter \cite{rudnerreview,oka2019floquet}. By subjecting a quantum system to a periodic external drive, it is possible to engineer effective Hamiltonians that exhibit nontrivial topological phases \cite{dabiri2021light,dabiri2021engineering,shafiei2024floquet,shafiei2025linearly}. Specially, time-periodicity can induce topological phases absent in the static case, including anomalous Floquet topological insulators \cite{rudner2013anomalous,lindner2011floquet} and topological pumps \cite{thoules1983quantization,minguzzi2022topological}. In these periodically driven systems, Floquet states and their associated quasienergies replace the conventional Bloch eigenstates and energy bands, providing a natural framework for extending the concepts of band topology and quantum geometry into the time domain.

Because the geometric structure of Floquet-Bloch states essentially encodes the nontrivial topological properties of Floquet systems, quantum geometry is a good framework for describing the static and dynamical aspects of periodically driven matter.

A central concept in modern condensed matter physics is the role of quantum geometry, embodied by the Berry curvature and the quantum metric, in characterizing Bloch bands and governing their linear and nonlinear responses \cite{ma2021topology,xiao2010berry,resta2011insulating,provost1980riemannian, PhysRevB.97.201117, bdhy-hnd2, yu2025quantum}. While the Berry curvature is intimately connected to Hall-type responses and topological invariants, the quantum metric governs wave-packet spreading, dielectric properties, and a variety of nonlinear optical phenomena \cite{yu2025quantum,orenstein2021topology,Neupert2013}. In static systems, these geometric quantities have been extensively investigated both theoretically and experimentally \cite{verma2026quantum, kang2025measurements}. Their extension to periodically driven (Floquet) systems, however, remains an active area of research, as the explicit time dependence of Floquet states introduces new conceptual and practical challenges.

Recently, several studies have explored the Berry curvature and quantum metric in periodically driven systems, revealing remarkable phenomena such as light-induced band geometry \cite{bao2022light,mciver2019light}, Floquet-engineered optical nonlinearities \cite{shan2021giant}, and nonlinear responses unique to Floquet bands \cite{dabiri2025dynamical,dabiri2025velocity}. However, a systematic treatment of time-dependent quantum geometry formulated directly in the Floquet-state basis—including its connections to optical conductivity, optical sum rules, and non-adiabatic topological charge pumping—has remained lacking. Moreover, the constraints imposed by discrete symmetries on the time-dependent quantum geometric tensor and its associated topological and integrated geometric quantities have yet to be systematically established.

In this work, we address these issues by defining a time-dependent quantum geometric tensor directly in the instantaneous Floquet-state basis and developing a unified framework for the time-dependent Berry curvature, $\Omega^{xy}(t)$, and quantum metric, $g^{xy}(t)$, of Floquet–Bloch states, where $x$ and $y$ denote spatial coordinates and $t$ is time. We derive a set of optical sum rules that relate the Fourier components of these geometric quantities to the optical conductivity and demonstrate that, under ideal Floquet-band occupations, the first-order DC Hall and longitudinal responses to a probe field at harmonic frequencies $n\Omega$ ($n\neq0$) vanish identically [Eqs.~(\ref{xzzero}) and (\ref{xxzero})]. These analytical predictions are verified numerically using the driven Rudner–Lindner–Berg–Levin (RLBL) model \cite{rudner2013anomalous}, as well as previously studied Floquet models \cite{seradjeh2020,dabiri2025dynamical}.

We also extend the concept of quantized charge pumping to the non-adiabatic regime by introducing a mixed Berry curvature, $\Omega_\alpha^{xt}$, involving both time and momentum derivatives. We show that the charge pumped during one driving period is equal to the sum of the Chern numbers of the occupied Floquet bands, thereby generalizing Thouless pumping \cite{thoules1983quantization} to periodically driven systems in which adiabaticity is not required. This intrinsic pumping mechanism contrasts with other non-adiabatic proposals that rely on external biases or carefully engineered driving protocols \cite{titum2016anomalous,malikis2022ideal}. Moreover, we identify the temporal component of the quantum metric, $g_\alpha^{tt}$, as the energy fluctuation of a Floquet band, while the mixed component, $g_\alpha^{xt}$, is interpreted as the correlation between polarization and energy.

In addition, we present a comprehensive symmetry analysis, demonstrating how time-reversal, sublattice (chiral), particle-hole, inversion, parity-time, rotational, and reflection symmetries constrain the time-dependent quantum geometric tensor and its associated topological invariants. The results, summarized in Table~\ref{symtab}, provide practical guidelines for engineering Floquet systems with tailored geometric properties and for identifying symmetry-protected zeros in the optical response and topological charge pumping.

Furthermore, we perform numerical simulations of both the RLBL model and a fully symmetric model [Eq.~(\ref{hsym})] to validate our analytical predictions. We compute the time-dependent quantum geometric quantities, verify the optical sum rules, demonstrate the emergence of topological edge states, and evaluate the optical conductivity under ideal Floquet-band occupations, where negative longitudinal conductivity, a characteristic signature of driven Floquet systems, is observed.

This paper is organized as follows. Sec.~\ref{prelim} and \ref{toposec} review essential Floquet formalism and topological invariants. In Sec.~\ref{geom} and \ref{metrsec} we define the time‑dependent Berry curvature and quantum metric, derive sum rules, and connect them to optical conductivities. Sec.~\ref{pump} discusses nonadiabatic quantized charge pumping and the interpretation of the quantum metric as fluctuations. Sec.~\ref{symsec} provides a detailed symmetry analysis. Numerical results on the RLBL and symmetric models are presented in Sec.~\ref{num}. We conclude in Sec.~\ref{conclusion}. Appendices contain proofs of gauge invariance and additional numerical verifications.

\section{Theory and Model}\label{prelim}
 
Our focus here is on the time-dependent quantum geometric tensor and its role in characterizing the geometry and dynamics of quantum states. To this end, we explore its manifestations through Berry curvature, quantum metric, optical sum rules, charge pumping, symmetry considerations, optical conductivity, energy fluctuations, and polarization–energy covariance, establishing the connections between quantum geometry and measurable dynamical responses. Therefore, we will describe how the optical, topological, transport, and symmetry features of Floquet systems are unified by a single geometric entity.

\subsection{Floquet theory}\label{prelim}
This subsection briefly reviews the notation and fundamental concepts of Floquet theory \cite{PhysRevLett.110.200403}. We consider a time-periodic Bloch Hamiltonian, $H(t)=H(t+T)$, where  $T=2\pi/\Omega$ is the driving period and $\Omega$ is the frequency of an external gauge vector. The corresponding Schr\"{o}dinger equation is
 
\begin{equation}
\begin{aligned}
(H(t)-i{{\partial }_{t}})|{{\psi }_{\alpha }}(t)\rangle =0 .
\end{aligned}
\label{schro1}
\end{equation}
where the momentum index $\mathbf{k}$ is suppressed for brevity (as it will be throughout the rest of the paper, unless there is a risk of confusion). According to the Floquet theorem, the eigenfunctions can be written as $|{{\psi }_{\alpha }}(t)\rangle=e^{-i \epsilon_\alpha t}|{{\phi }_{\alpha }}(t)\rangle$, where the Greek index $\alpha$ labels the band, $|{{\phi }_{\alpha }}(t)\rangle=|{{\phi }_{\alpha }}(t+T)\rangle$ are the \emph{Floquet states}, and  $\epsilon_\alpha$  are the \emph{quasienergies}, defined modulo  $\Omega$. The Floquet Brillouin zone (FBZ) is defined by restricting quasienergies to $-\Omega/2<\epsilon<\Omega/2$ and momentum to the regular Brillouin zone. The bands can then be ordered according to their quasienergy values in the FBZ and assigned an $\alpha$ index.

Inserting the Floquet ansatz into the Schr\" {o}dinger equation (\ref{schro1}) gives the following evolution equation for the Floquet states: 
\begin{equation}
\begin{aligned}
(H(t)-i{{\partial }_{t}})|{{\phi }_{\alpha }}(t)\rangle ={{\epsilon }_{\alpha }}|{{\phi }_{\alpha }}(t)\rangle .
\end{aligned}
\label{schro2}
\end{equation}
Substituting the Fourier expansions  ${{H}}(t)={{e}^{-in\Omega t}}{{H}^{(n)}}$ and $|{\phi }_{\alpha }(t)\rangle ={{e}^{-im\Omega t}}|{\phi }_{\alpha }^{(m)}\rangle $ for the time-periodic Hamiltonian and Floquet states into (\ref{schro2}), one obtains
\begin{equation}
\begin{aligned}
{{\sum }_{n}}{{H}^{(m-n)}}|\phi _{\alpha }^{(n)}\rangle -m\Omega |\phi _{\alpha }^{(m)}\rangle ={{\epsilon }_{\alpha }}|\phi _{\alpha }^{(m)}\rangle .
\end{aligned}
\label{hmn}
\end{equation}
This equation represents an eigenvalue problem involving a static, infinite-dimensional Hamiltonian. In numerical computations, the Hamiltonian should be truncated to a dimension that guarantees convergent results (see also \cite{dabiri2025dynamical}). It can also be shown that an orthogonality relation holds for the Floquet states, given by
\begin{equation}
\begin{aligned}
\langle {{\phi }_{\alpha }}(t)|{{\phi }_{\beta }}(t)\rangle ={{\delta }_{\alpha \beta }}.
\end{aligned}
\label{orthoeq}
\end{equation}
which we assume hereafter.

In terms of the Floquet states, the time evolution operator from $t=0$ to $t$ takes the form
\begin{equation}
\begin{aligned}
{{U}_{t}}=\sum\limits_{\alpha }{{{U}_{t}}|{{\phi }_{\alpha }}(0)\rangle \langle {{\phi }_{\alpha }}(0)|}=\sum\limits_{\alpha }{{{e}^{-i{{\epsilon }_{\alpha }}t}}|{{\phi }_{\alpha }}(t)\rangle \langle {{\phi }_{\alpha }}(0)|}.
\end{aligned}
\label{ }
\end{equation}
The Floquet states and quasienergies are determined by the eigenvalue equation of the evolution operator over a single time period:
\begin{equation}
\begin{aligned}
{{U}_{T}}|{{\phi }_{\alpha }}(0)\rangle ={{e}^{-i{{\epsilon }_{\alpha }}T}}|{{\phi }_{\alpha }}(0)\rangle. 
\end{aligned}
\label{UTeq}
\end{equation}
We hence proceed by defining the effective static Hamiltonian as
\begin{equation}
\begin{aligned}
H_{\text{eff}}^{\mu }=&\frac{i}{T}\sum\limits_{\alpha }{{{\ln }_{\mu }}({{e}^{-i{{\epsilon }_{\alpha }}T}})|{{\phi }_{\alpha }}(0)\rangle \langle {{\phi }_{\alpha }}(0)|}\\
&=\sum\limits_{\alpha }{\epsilon _{\alpha }^{\mu }|{{\phi }_{\alpha }}(0)\rangle \langle {{\phi }_{\alpha }}(0)|},
\end{aligned}
\label{ }
\end{equation}
where $\mu$ denotes the branch cut of logarithm i.e. $\mu -\Omega<\epsilon _{\alpha }^{\mu }<\mu $.

By diagonalizing the time-evolution operator at any given time, one obtains the \emph{phase bands} $\varphi(t)$ and \emph{phase states}  $|{{\chi }_{\alpha }}(t)\rangle $, given by
\begin{equation}
\begin{aligned}
{{U}_{t}}=\sum\limits_{\alpha }{{{e}^{-i{{\varphi }_{\alpha }}(t)}}|{{\chi }_{\alpha }}(t)\rangle \langle {{\chi }_{\alpha }}(t)|},
\end{aligned}
\label{utdef}
\end{equation}
forming a complete basis. Another useful quantity is the periodized evolution operator, defined as
\begin{equation}
\begin{aligned}
U_{t}^{\mu }={{U}_{t}}{{e}^{iH_{\text{eff}}^{\mu }t}}=\sum\limits_{\alpha }{{{e}^{i(\epsilon _{\alpha }^{\mu }-{{\epsilon }_{\alpha }})t}}|{{\phi }_{\alpha }}(t)\rangle \langle {{\phi }_{\alpha }}(0)|}.
\end{aligned}
\label{umu}
\end{equation}
It therefore follows that $U_{0}^{\mu }=U_{T}^{\mu }$.  

\subsection{topological aspects}\label{toposec}
In the remainder of this paper, we focus on two-dimensional (2D) Floquet systems with gapped quasienergy bands. The topological analysis presented here is restricted to Floquet systems without internal symmetries, while its extension to other symmetry classes can be performed straightforwardly. Note that since $|{{\phi }_{\alpha }}(0)\rangle$  and $|{{\phi }_{\alpha }}(t)\rangle$ are related by a smooth unitary evolution, they yield the same instantaneous Chern number. Consequently, the Chern number of the Floquet states does not capture the full topology of a time-periodic system, but rather provides the topological invariant associated with a single-cycle evolution. An additional topological invariant is therefore required to fully characterize the topological properties of 2D Floquet systems.
 
During the time evolution, the phase bands may exhibit gap closings, referred to as phase band singularities. As pointed out by Nathan and Rudner \cite{nathan2015topological}, gap closings of the phase bands at the Floquet zone edge $\epsilon=\Omega/2$ can be topological (i.e., they cannot be removed by small perturbations to the Hamiltonian), whereas gap closings at zero quasienergy $\epsilon=0$ are no longer topological and can be trivialized by smoothly deforming the phase bands.

The full topological behavior of the system can be obtained straightforwardly from the phase bands and phase states. The number of edge states in the quasienergy gap  $\mu$ is given by the winding number $W_3(\mu)$, defined as
\begin{equation}
\begin{aligned}
{{W}_{3}}(\mu )=\sum\limits_{{{\epsilon }_{n}}<\mu }{{{C}_{n}^{xy}}}-\sum\limits_{i}{q_{i}^{ZES}},
\end{aligned}
\label{w3eq}
\end{equation}
where $ZES$ states zero-energy states. In this expression, the first term represents the sum of the Chern numbers of all Floquet bands below the gap, and the second term represents the sum of the topological charges of the gap-closing points at the FBZ edge,  $\epsilon=\Omega/2$. It should be noted that at $t=T$, the phase states reduce to the Floquet states, and the phase band energies correspond to the quasienergies: $|\chi_\alpha(T)\rangle=|\phi_\alpha(T)\rangle,~~\varphi_\alpha(T)=\epsilon_\alpha T$.
 
The topological charge at a gap-closing point $t=t_0, \epsilon=\Omega/2$ can also be obtained by linearizing the matrix elements of the evolution operator \cite{nathan2015topological}. In terms of the periodized evolution operator given in Eq.~(\ref{umu}), the topological invariant of a 2D Floquet system with no internal symmetries admits a more formal definition as the winding number associated with a specific gap $\mu$:
\begin{equation}
\begin{aligned}
W_3({\mu })=\int\limits_{\mathbf{k}}^{\,}{\int\limits_{0}^{T}{\frac{dt}{6}\text{Tr}\{U_{t}^{{{\mu }^{-1}}}{{\partial }_{t}}U_{t}^{\mu }[U_{t}^{{{\mu }^{-1}}}{{\partial }_{{{k }_{x}}}}U_{t}^{\mu },U_{t}^{{{\mu }^{-1}}}{{\partial }_{{{k }_{z}}}}U_{t}^{\mu }]\}}}.
\end{aligned}
\label{ }
\end{equation}
where the integral runs over the Brillouin zone, $\int_{\mathbf{k}}=\sum_{\mathbf{k}}=\int_{BZ}{\frac{{{d}^{2}}\mathbf{k}}{{{(2\pi )}^{2}}}}$. The Chern number of a given Floquet band is given by the difference between the winding numbers of the adjacent gaps above and below it.

\subsection{time dependent Berry curvature, sum rules and quantized conductivities}\label{geom}

For periodically driven systems, several geometric formulations have been developed, including approaches based on instantaneous Bloch bands, extended (Sambe) Hilbert-space eigenstates, and Berry curvatures derived from effective Floquet Hamiltonians \cite{7l91-gw77,b3pw-my97}. These methods have proven highly effective in describing time-averaged transport properties and topological phases. However, they do not directly capture the instantaneous geometric evolution of physical Floquet states throughout a driving cycle. In particular, the Sambe-space formulation \cite{PhysRevA.7.2203} maps the periodically driven problem onto a static eigenvalue problem in an enlarged Hilbert space, making the physical interpretation of the resulting geometric quantities less transparent. In contrast, the effective Floquet Hamiltonian describes only the stroboscopic evolution over one driving period and therefore omits information about the intra-period micromotion \cite{asteria2026micromotion}.

Motivated by these considerations, we explicitly define the quantum geometric tensor in the basis of time-periodic Floquet states. This formulation continuously reduces to the conventional quantum geometry of static Bloch bands in the zero-frequency limit, provides a direct connection to experimentally accessible time-resolved optical responses, and retains the full time dependence associated with micromotion. Consequently, it offers a complementary framework for Floquet quantum geometry that is particularly suited for investigating optical phenomena, dynamical transport, and nonadiabatic topological effects.

In the basis of Floquet states, the time-dependent Berry curvature, in $({\bf {r}}, t)$ manifold, is defined as 
\begin{equation}
\begin{aligned}
{\Omega} _{\beta }^{xy}(t)&=2\sum\limits_{\alpha}{\operatorname{Im}(\langle {{\phi }_{\beta }}(t)|{{x}^{e}}|{{\phi }_{\alpha }}(t)\rangle \langle {{\phi }_{\alpha }}(t)|{{y}^{e}}|{{\phi }_{\beta }}(t)\rangle )}\\
&\equiv \sum\limits_{\alpha}{\Omega} _{\beta \alpha}^{xy}(t).
\end{aligned}
\label{omt}
\end{equation}
where $\mathbf{r}^e=(x^e,y^e,z^e)$ denotes the interband position operator, with matrix elements between Floquet states defined as  $\mathbf{r}_{\beta \alpha }^{e}\equiv \langle {{\phi }_{\beta }}(t)|{\mathbf{r}^{e}}|{{\phi }_{\alpha }}(t)\rangle=(1-\delta_{\alpha \beta})\langle {{\phi }_{\beta }}(t)|i\partial_\mathbf{k}|{{\phi }_{\alpha }}(t)\rangle$ \cite{shankar2017quantum} and $ {\Omega} _{\beta \alpha}^{xy}(t)$ is band-resolved Berry curvature.

The Chern number of a Floquet band follows from integrating the Berry curvature over the Brillouin zone:
\begin{equation}
\begin{aligned}
C_\beta^{xy}=C_\beta^{xy}(t)=2\pi\sum\limits_{\mathbf{k}}{\Omega} _{\beta }^{xy}(t).
\end{aligned}
\label{ }
\end{equation}
It should be noted that the integral of ${\Omega}_\beta^{xy} (t)$ over the Brillouin zone is time-independent.
 
As stated earlier, the Floquet states $|\phi_\alpha(0)\rangle$ and $|\phi_\alpha(t)\rangle$ yield the same time-independent Chern number. Defining $\mathbf{r}_{\beta \alpha }^{e(n)}\equiv \int_{0}^{T}{\frac{dt}{T}{{e}^{in\Omega t}}}\langle {{\phi }_{\beta }}(t)|{\mathbf{r}^{e}}|{{\phi }_{\alpha }}(t)\rangle $, one can exploit this fact to obtain a sum rule:
\begin{equation}
\begin{aligned}
&{\int_{0}^{T}{\frac{dt{{e}^{in\Omega t}}}{T}}}\sum\limits_{\mathbf{k}\alpha }{2\operatorname{Im}\{x_{\beta \alpha }^{e}}y_{\alpha \beta }^{e}\}=0,\,\,\text{for}\,n\ne 0, \\ 
 & \Rightarrow \,\,\,\,\, \,\,\,\sum\limits_{\mathbf{k}\alpha j}x_{\beta \alpha }^{e(j)}y_{\alpha \beta }^{e(-j+n)}-x_{ \alpha \beta}^{e(j)}y_{ \beta \alpha}^{e(-j+n)}=0,\,\,\text{for}\,n\neq 0.
\end{aligned}
\label{sumr}
\end{equation}
This has important physical consequences. As shown in \cite{dabiri2025dynamical,dabiri2025velocity}, the optical conductivity of a Floquet system is time periodic, taking the form $\sigma (\omega )=\sum\limits_{n\in \mathbb{Z}}{{{e}^{-in\Omega t}}}\sigma _{\,}^{(n)}(\omega )$. Consequently, the first-order interband optical conductivity is given by
\begin{equation}
\begin{aligned}
   \sigma _{xy}^{[1]e(n)}(\omega )&=\sum\limits_{j\alpha \beta \mathbf{k}}{{{f}_{\beta \alpha }}\frac{v_{\beta \alpha }^{x(j+n)}y_{\alpha \beta }^{e(-j)}}{{{\epsilon }_{\alpha \beta }}+j\Omega -\omega }} \\ 
 & =\sum\limits_{j\alpha \beta \mathbf{k}}{i{{f}_{\beta \alpha }}\frac{({{\epsilon }_{\beta \alpha }}-(j+n)\Omega )x_{\beta \alpha }^{e(j+n)}y_{\alpha \beta }^{e(-j)}}{{{\epsilon }_{\alpha \beta }}+j\Omega -\omega }} 
\end{aligned}
\label{oc1}
\end{equation}
where $f_{\beta\alpha}=f_{\beta}-f_{\alpha}$ and $f_{\beta}$ denotes the occupation of Floquet band $\beta$. This occupation is not generally described by the Fermi–Dirac distribution; instead, it depends on several factors, including the driving protocol during the switch-on process and the nature of dissipation mechanisms. The velocity operator is given by $v^x=\partial_{k_x}H(t)$, with $j\in\mathbb{Z}$, and the velocity and position matrix elements satisfy the relation
$
i \mathbf{r}^{e(-j)}_{\alpha\neq\beta}(\epsilon_{\alpha\beta}+j\Omega)=\mathbf{v}_{\alpha\beta}^{(-j)}.
$
It should be noted that this relation is generally modified when time-dependent phase factors, corresponding to a time-dependent gauge choice, are introduced.
The reality of conductivity imposes the following equality \cite{seradjeh2020}:
\begin{equation}
\begin{aligned}
\sigma^{(n)}(\omega)=\sigma^{(-n)}(-\omega)^*
\end{aligned}
\label{reals}
\end{equation}
By combining Eq.~(\ref{oc1}) with the assumption of ideal occupation—i.e., Floquet bands that are either fully occupied or empty—the sum rule of Eq.~(\ref{sumr}) takes the form
\begin{equation}
\begin{aligned}
  & \sigma _{xy}^{[1]e(-n)}(n\Omega ) =-i\sum\limits_{j\alpha \beta \mathbf{k}}{{{f}_{\beta \alpha }}x_{\beta \alpha }^{e(j-n)}y_{\alpha \beta }^{e(-j)}}\text{ } \\ 
 & =-i\sum\limits_{j\alpha \beta \mathbf{k}}{{{f}_{\beta }}(x_{\beta \alpha }^{e(j-n)}y_{\alpha \beta }^{e(-j)}-y_{\beta \alpha }^{e(-j)}x_{\alpha \beta }^{e(j-n)})} \\ 
 & =\sum\limits_{\beta \mathbf{k}}{{{f}_{\beta }}\Omega _{\beta }^{xy(-n)}=0} 
,\,\,\text{for}\,\,n\ne 0~\&~\text{ideal occupation}.
\end{aligned}
\label{xzzero}
\end{equation}
Equation~(\ref{xzzero}) shows that the first-order DC Hall response to a probe of frequency $n\Omega$ acting on a time-periodic system with fundamental frequency  $\Omega$ is identically zero. Note that $\sigma _{xy}^{[1]e(0)}(0)$ is also quantized, taking the value of the sum of the Chern numbers of all filled bands.

An analogous calculation applies to the longitudinal optical conductivity:
\begin{equation}
\begin{aligned}
  & \sigma _{xx}^{[1]e(-n)}(n\Omega ) =-i\sum\limits_{j\alpha \beta \mathbf{k}}{{{f}_{\beta \alpha }}x_{\beta \alpha }^{e(j-n)}x_{\alpha \beta }^{e(-j)}}\text{ } \\ 
 & =-i\sum\limits_{j\alpha \beta \mathbf{k}}{{{f}_{\beta }}(x_{\beta \alpha }^{e(j-n)}x_{\alpha \beta }^{e(-j)}-x_{\beta \alpha }^{e(-j)}x_{\alpha \beta }^{e(j-n)})} \\ 
 & =-i\sum\limits_{j\alpha \beta \mathbf{k}}{{{f}_{\beta }}(x_{\beta \alpha }^{e(j-n)}x_{\alpha \beta }^{e(-j)}-x_{\beta \alpha }^{e(j-n)}x_{\alpha \beta }^{e(-j)})} =0. \\ 
\end{aligned}
\label{xxzero}
\end{equation}
In obtaining the third line, we have used the fact that the index $j$ runs over $\mathbb{Z}$, allowing the substitution $j-n\rightarrow -j^\prime$, so that $-j \rightarrow j^\prime-n$. In conclusion, if the Floquet bands are filled or empty, the intraband response also vanishes, and the first-order DC longitudinal response of a Floquet system to a probe at frequency $n\Omega$ is zero. Equations~(\ref{xzzero}) and (\ref{xxzero}) constitute the two main results of this paper. Physically, they claim that the system cannot absorb or emit energy from a probe field at multiples of the driving frequency when Floquet bands are fully filled or empty without altering its occupation; the net DC response at harmonics disappears because the upward and downward photon-assisted transitions between Floquet bands precisely cancel each other out.
They can also be verified experimentally on suitable platforms, offering a novel perspective on the optical response of Floquet systems. A comprehensive numerical verification of the identities (\ref{xzzero}) and (\ref{xxzero}) introduced in this work can be found in \cite{seradjeh2020}. Specifically, Figs.~6, 7, and 10 of Ref.~\cite{seradjeh2020} show that $\sigma^{[1]e(-1)}_{xx}(\Omega)$ in general, and $\sigma^{[1]e(-1)}_{xy}(\Omega)$ for ideal occupation, vanish for the driven Haldane model and the driven quantum well. Interestingly, Fig.~10(d) of \cite{seradjeh2020} shows that  $\sigma^{[1]e(-1)}_{xy}(\Omega)$ is zero only for ideal occupation, not for the quench occupation of Floquet states. Furthermore, Figs.~1, 2, and 3 of Ref.~\cite{dabiri2025dynamical} confirm Eqs.~(\ref{xzzero}) and (\ref{xxzero}) for the driven Su-Schrieffer-Heeger model. We also provide numerical confirmation of these equations for the driven RLBL model up to the fifth Fourier component in Fig.~\ref{ocfig}.

\subsection{Time dependent quantum metric}\label{metrsec}
It is well established that the Berry curvature quantifies the phase difference between adjacent wave functions in the Brillouin zone, whereas the quantum metric captures the distance between neighboring wave functions in parameter space. If the wave functions depend on a parameter vector $\boldsymbol{\lambda}$ and obey the normalization condition $\langle\psi(\boldsymbol{\lambda})|\psi(\boldsymbol{\lambda})\rangle=1$, then
\begin{equation}
\begin{aligned}
1-|\langle\psi(\boldsymbol{\lambda})|\psi(\boldsymbol{\lambda}+d\boldsymbol{\lambda})\rangle|^2=\sum_{\mu \nu}g^{\mu \nu}d\boldsymbol{\lambda_\mu}d\boldsymbol{\lambda_\nu},
\end{aligned}
\label{}
\end{equation}
where $g^{\mu \nu}$ is the quantum metric.

This concept can be generalized to Floquet systems by defining the time-periodic quantum metric in the Floquet-state basis as
\begin{equation}
\begin{aligned}
{g} _{\beta }^{xy}(t)&= \sum\limits_{\alpha}{\operatorname{Re}(\langle {{\phi }_{\beta }}(t)|{{x}^{e}}|{{\phi }_{\alpha }}(t)\rangle \langle {{\phi }_{\alpha }}(t)|{{y}^{e}}|{{\phi }_{\beta }}(t)\rangle )}\\
&\equiv\sum\limits_{\alpha}{g} _{\beta \alpha}^{xy}(t)
\end{aligned}
\label{gt}
\end{equation}
The definitions in Eqs.~(\ref{omt}) and (\ref{gt}) imply that
\begin{equation}
\begin{aligned}
&{g} _{\beta }^{mn}(t)={g} _{\beta }^{nm}(t),~~~{\Omega} _{\beta }^{mn}(t)=-{\Omega} _{\beta }^{nm}(t). 
\end{aligned}
\label{}
\end{equation}
Hence, the Berry curvature and the quantum metric are the imaginary (anti-symmetric) and real (symmetric) components of a more general quantity, the quantum geometric tensor, which can be written as $\mathcal{G}^{xy}_\beta(t)=g^{xy}_\beta(t)-\frac{i}{2}{\Omega}^{xy}_\beta(t)$.

The time-averaged quantum metric and Berry curvature take the form
\begin{eqnarray}
&g_{\beta }^{xy(0)}=\sum\limits_{\alpha j}{\operatorname{Re}(x_{\beta \alpha }^{e(j)}}y_{\alpha \beta }^{e(-j)})\equiv \sum\limits_{j}{g_{\beta }^{xy\{j\}}}\equiv \sum\limits_{\alpha j }{g_{\beta \alpha}^{xy\{j\}}},\\
&\Omega_{\beta }^{xy(0)}=2\sum\limits_{\alpha j}{\operatorname{Im}(x_{\beta \alpha }^{e(j)}}y_{\alpha \beta }^{e(-j)})\equiv \sum\limits_{j}{\Omega_{\beta }^{xy\{j\}}}\equiv \sum\limits_{\alpha j }{\Omega_{\beta \alpha}^{xy\{j\}}},
\label{ }
\end{eqnarray} 
while the time-averaged quantum geometric tensor is
\begin{equation}
\begin{aligned}
\mathcal{G}_{\beta }^{xy(0)}=\sum\limits_{\alpha j}{(x_{\beta \alpha }^{e(j)}}y_{\alpha \beta }^{e(-j)})\equiv \sum\limits_{j}{\mathcal{G}_{\beta }^{xy\{j\}}}\equiv \sum\limits_{\alpha j }{\mathcal{G}_{\beta \alpha}^{xy\{j\}}}.
\end{aligned}
\label{ }
\end{equation} 


Combining Eqs.~(\ref{oc1}) and (\ref{reals}) yields
\begin{equation}
\begin{aligned}
  & \sigma _{xy}^{(n)}(\omega )+\sigma _{xy}^{(-n)}(-\omega ) \\ 
 & =\sum\limits_{j\alpha \beta \mathbf{k}}{{{f}_{\beta \alpha }}\frac{-2(\epsilon_{\beta\alpha}-(j+n)\Omega)\operatorname{Im}[x_{\beta \alpha }^{e(j+n)}y_{\alpha \beta }^{e(-j)}]}{{{\epsilon }_{\alpha \beta }}+j\Omega -\omega }}, \\
\end{aligned}
\label{2sigma1}
\end{equation}
moreover,
 \begin{equation}
\begin{aligned}
& \sigma _{xy}^{(n)}(\omega )-\sigma _{xy}^{(-n)}(-\omega ) \\ 
 & =\sum\limits_{j\alpha \beta \mathbf{k}}{{{f}_{\beta \alpha }}\frac{2i(\epsilon_{\beta\alpha}-(j+n)\Omega)\operatorname{Re}[x_{\beta \alpha }^{e(j+n)}y_{\alpha \beta }^{e(-j)}]}{{{\epsilon }_{\alpha \beta }}+j\Omega -\omega }}, 
\end{aligned}
\label{2sigma}
\end{equation}
Equation~(\ref{2sigma1}) implies  $ \operatorname{Im}\{\sigma _{xy}^{(n)}(\omega )+\sigma _{xy}^{(-n)}(-\omega )\} =2\pi (\omega +n\Omega )\sum\limits_{j\alpha \beta \mathbf{k}}{{{f}_{\beta \alpha }}\operatorname{Im}[x_{\beta \alpha }^{e(j+n)}y_{\alpha \beta }^{e(-j)}]}\delta ({{\epsilon }_{\alpha \beta }}+j\Omega -\omega )
$ and Eq.~(\ref{2sigma}) leads to  $   \operatorname{Re}\{\sigma _{xy}^{(n)}(\omega )-\sigma _{xy}^{(-n)}(-\omega )\} =2\pi (\omega +n\Omega )\sum\limits_{j\alpha \beta \mathbf{k}}{{{f}_{\beta \alpha }}\operatorname{Re}[x_{\beta \alpha }^{e(j+n)}y_{\alpha \beta }^{e(-j)}]}\delta ({{\epsilon }_{\alpha \beta }}+j\Omega -\omega )
$. Consequently, the following sum rules follow
 \begin{equation}
\begin{aligned}
 &  \int_{-\infty }^{\infty }d\omega{\frac{\operatorname{Re}\{\sigma _{xy}^{(n)}(\omega )-\sigma _{xy}^{(-n)}(-\omega )\}}{2\pi (\omega +n\Omega )}} \\ 
 &~~~~~~~~~~~~~~~~~~~~~~~~~= \sum\limits_{j\alpha \beta \mathbf{k}}{{{f}_{\beta \alpha }}\operatorname{Re}[x_{\beta \alpha }^{e(j+n)}y_{\alpha \beta }^{e(-j)}]},
\end{aligned}
\label{resms}
\end{equation}
and also
 \begin{equation}
\begin{aligned}
  & \int_{-\infty }^{\infty }d\omega{\frac{\operatorname{Im}\{\sigma _{xy}^{(n)}(\omega )+\sigma _{xy}^{(-n)}(-\omega )\}}{2\pi (\omega +n\Omega )}} \\ 
 &~~~~~~~~~~~~~~~~~~~~~~~~~= \sum\limits_{j\alpha \beta \mathbf{k}}{{{f}_{\beta \alpha }}\operatorname{Im}[x_{\beta \alpha }^{e(j+n)}y_{\alpha \beta }^{e(-j)}]}.
\end{aligned}
\label{imsps}
\end{equation}
Equations~(\ref{resms}) and (\ref{imsps}) demonstrate that the quantum geometric tensor components expressed in the Floquet-state basis (including Eqs.~(\ref{omt}) and (\ref{gt})) fully determine the optical conductivity, thereby providing a pathway for their experimental measurement.

Let us consider the dissipative part of the optical conductivity, consider elliptically polarized light described by an angle $0<\vartheta<\pi$ corresponding to right-handed and  $-\pi<\vartheta<0$  to left-handed polarization. The zeroth Fourier component of the optical conductivity reads
\begin{equation}
\begin{aligned}
  \operatorname{Re}\sigma _{\vartheta}^{[1]e(0)}(\omega )=&\pi \omega\sum\limits_{\mathbf{k}\alpha \beta j}{{{{f}_{\beta \alpha }}{G}_{\beta \alpha}^{\{j\}}(\vartheta)\delta ({{\epsilon }_{\alpha \beta }}+j\Omega -\omega )}}. 
\end{aligned}
\label{sigvar}
\end{equation}
where
\begin{equation}
\begin{aligned}
{G}_{\beta \alpha}^{\{j\}}(\vartheta)={g_{\beta \alpha}^{xx\{j\}}} \cos^2 \vartheta+{g_{\beta \alpha}^{yy\{j\}}} \sin^2 \vartheta+ {\Omega_{\beta \alpha}^{xy\{j\}}} \sin\vartheta \cos\vartheta. 
\end{aligned}
\label{}
\end{equation}
Here $\vartheta=\pi/4 (-\pi/4)$ corresponds to right (left) circularly polarized light. Circular dichroism—the differential absorption of right- and left-handed circularly polarized light—offers a direct probe of the quantum metric and curvature components. Equation~(\ref{sigvar}) indicates a practical route to extract the components of the quantum geometric tensor and, in particular, their $\mathbf{k}$ dependence, which is especially tractable for two-band systems where only two band-index terms, separable in experiments, contribute.

The quantum metric also appears in nonlinear optical responses. The second-order response, for instance, is nonzero only in noncentrosymmetric systems. As shown in Eq.~(E7) of Ref.~\cite{dabiri2025dynamical}, the dominant contribution to the injection current—a DC bulk photovoltaic current arising at second order in the probe field—for $\omega\neq n\Omega/2,$ with $n\in \mathbb{Z}$, takes the form
\begin{equation}
\begin{aligned}
  \sigma _{xyz}^{\text{inj}}& ={-\pi }\tau\sum\limits_{\alpha \beta j\mathbf{k}}{\Delta _{\alpha \beta }^{x(0)}{{f}_{\alpha \beta }}}\mathcal{G}_{\alpha \beta }^{yz\{j\}}\delta (\omega -{{\epsilon }_{\beta \alpha }}+j\Omega ) \\ 
 & ={-\pi }{\tau }\sum\limits_{\alpha \beta j\mathbf{k}}{\Delta _{\alpha \beta }^{x(0)}{{f}_{\alpha \beta }}}y_{\alpha \beta }^{e(j)}z_{\beta \alpha }^{e(-j)}\delta (\omega -{{\epsilon }_{\beta \alpha }}+j\Omega ),
\end{aligned}
\label{inj0}
\end{equation}
where $\tau$ is the relaxation time and $\Delta _{\alpha \beta }^{x(0)}=v _{\alpha \alpha }^{x(0)}-v _{\beta \beta }^{x(0)}$. The injection current given by Eq.~(\ref{inj0}) originates from the time-averaged velocity difference between the valence and conduction bands and is a purely interband effect. A key difference between static and Floquet systems is that, while the injection current always vanishes at subgap frequencies in the static case, it may be nonzero in Floquet systems owing to an extra resonance at  $\omega=\Omega-\epsilon_{\alpha\beta}$  in Eq.~(\ref{inj0}), which can occur below the quasienergy gap.

We next examine another second-order response, the Berry curvature dipole term. When two probe fields at frequencies  $\omega$ and $-\omega$ are applied, the interband–intraband contribution reads
\begin{equation}
\begin{aligned}
\sigma _{xyz}^{[2]ei(0)}=\frac{i}{2\omega }\sum\limits_{\beta \mathbf{k}}{{{f}_{\beta }}[{{\partial }_{{{k}_{z}}}}\Omega _{\beta }^{xy(0)}-{{\partial }_{{{k}_{y}}}}\Omega _{\beta }^{xz(0)}}].
\end{aligned}
\label{berdipole}
\end{equation}
Unlike the injection current, the Berry curvature dipole scales inversely with the probe frequency rather than the relaxation time. This photocurrent is nonzero only for partially filled bands and is absent in ideal insulators. Under parity-time symmetry (PTS), the Berry curvature vanishes, and consequently Eq.~(\ref{berdipole}) evaluates to zero.

\subsection{Non-adiabatic quantized charge pumping and energy-polarization fluctuation}\label{pump}
In Eq.~(\ref{omt}), we introduced the time-dependent Berry curvature formulated in the Floquet-state basis. A natural question is how this construction is modified when the momentum derivative is replaced by a time derivative. To address this question, we evaluate the expectation value of the position operator along the $x$ direction for a Floquet state:
\begin{equation}
\begin{aligned}
  & \langle x\rangle_\alpha (t)=\langle {{\psi }_{\alpha }}(t)|x|{{\psi }_{\alpha }}(t)\rangle=i\langle {{\psi }_{\alpha }}(t)|{{\partial }_{{{k}_{x}}}}|{{\psi }_{\alpha }}(t)\rangle.  \\ 
\end{aligned}
\label{ }
\end{equation}
Although this expectation value is gauge-dependent, its integral over $k_x$—the \emph{polarization} along  $x$—is a gauge-invariant quantity of primary interest. The net displacement of the charge center after a single period of evolution reads
\begin{equation}
\begin{aligned}
 C^{xt}_\alpha&=\iint{dt}\frac{dk_x}{2\pi }\Omega _{\alpha }^{xt}, \\ 
  \Omega _{\alpha }^{xt}&=i\langle {{\partial }_{t}}{{\psi }_{\alpha }}(t)|{{\partial }_{{{k}_{x}}}}{{\psi }_{\alpha }}(t)\rangle -i\langle {{\partial }_{k_x}}{{\psi }_{\alpha }}(t)|{{\partial }_{t}}{{\psi }_{\alpha }}(t)\rangle  \\ 
 & =i\langle {{\partial }_{t}}{{\phi }_{\alpha }}(t)|{{\partial }_{{{k}_{x}}}}{{\phi }_{\alpha }}(t)\rangle -i\langle {{\partial }_{{{k}_{x}}}}{{\phi }_{\alpha }}(t)|{{\partial }_{t}}{{\phi }_{\alpha }}(t)\rangle  \\ 
 & =2\sum\limits_{\beta \neq \alpha }{\operatorname{Im}\langle {{\phi }_{\alpha }}(t)|{{\partial }_{t}}{{\phi }_{\beta }}(t)\rangle \langle {{\phi }_{\beta }}(t)|{{\partial }_{{{k}_{x}}}}{{\phi }_{\alpha }}(t)\rangle }.
\end{aligned}
\label{pumpx}
\end{equation}
The proof of this formula is presented in Sec.~\ref{apppump} of appendix. We emphasize that the mixed Berry curvature, $\Omega_{\alpha}^{xt}$, remains gauge invariant even when the Bloch functions acquire a time-dependent phase. Therefore, for ideal Floquet-band occupations (i.e., completely filled or empty bands), the charge pumped during a single driving cycle is equal to the sum of the Chern numbers of the occupied Floquet bands and is quantized. This result generalizes the original adiabatic Thouless charge pumping mechanism \cite{thoules1983quantization} to the non-adiabatic regime. The essential requirements are non-degenerate, gapped quasienergy bands with ideal occupation, rather than adiabatic evolution.

Although achieving ideal occupation of Floquet states remains experimentally challenging, several proposals have been developed to control Floquet-state populations on demand. These include specially designed switch-on protocols based on optimal control theory involving multiple driving frequencies \cite{optimal2022}, as well as coupling the system to engineered Fermi and Bose baths \cite{seeth}, thereby providing potential routes toward the observation of non-adiabatic charge pumping.

We note that, although the pumped charge in Eq.~(\ref{pumpx}) contains an explicit dependence on $k_y$, it is independent of any particular choice of $k_y$ because the Floquet bands are assumed to be non-degenerate. Consequently, the Floquet states are related by a smooth unitary evolution, which preserves the corresponding Chern number. In the adiabatic limit, the temporal variation of the Floquet states becomes sufficiently slow that the contribution of $i\partial_t$ in Eq.~(\ref{schro2}) can be neglected. The Floquet states then reduce to the instantaneous eigenstates of the time-dependent Hamiltonian. Therefore, our framework naturally includes adiabatic quantized Thouless pumping as a limiting case in the low-frequency regime.

Charge pumping in Floquet–Bloch bands has recently been observed in ultracold fermionic systems in optical lattices using a two-frequency drive \cite{minguzzi2022topological}. We emphasize that our proposal differs fundamentally from this approach. In their work, after the Floquet states are prepared, the parameters of the two-frequency drive are varied adiabatically to realize quantized charge pumping. In contrast, the charge pumping mechanism presented here occurs intrinsically after each driving period, arising from the nontrivial Berry curvature defined in Eq.~(\ref{pumpx}), without requiring additional adiabatic parameter modulation.

Another proposal for non-adiabatic quantized charge pumping was introduced in Ref.~\cite{titum2016anomalous}, where anomalous edge states enable pumping despite the Floquet bands having zero Chern numbers. However, this mechanism differs from ours in that it does not produce intrinsic charge pumping; instead, it requires an additional bias or asymmetric occupation of edge states. Furthermore, the approach proposed in Ref.~\cite{malikis2022ideal} relies on a highly constrained and finely tuned driving protocol, whereas our framework provides a more general mechanism for non-adiabatic quantized charge pumping.

A further quantity of interest is the mean energy of a Floquet band, expressed as
\begin{equation}
\begin{aligned}
  & \frac{1}{T}\int_0^T dt{{\langle E\rangle }_{\alpha }}(t)=\frac{1}{T}\int_0^T dt{\langle {{\psi }_{\alpha }}(t)|H(t)|{{\psi }_{\alpha }}(t)\rangle }\\
&=\frac{1}{T}\int_0^T dt{\langle {{\psi }_{\alpha }}(t)|i{{\partial }_{t}}|{{\psi }_{\alpha }}(t)\rangle }  \\ 
 & =\frac{1}{T}\int_0^T dt{\langle \phi _{\alpha }^{(m)}|({{\epsilon }_{\alpha }}+n\Omega )|\phi _{\alpha }^{(n)}\rangle }{{e}^{i(m-n)\Omega t}}\\
 &={\epsilon }_{\alpha }+\sum_n n\Omega \langle \phi _{\alpha }^{(n)}|\phi _{\alpha }^{(n)}\rangle={\epsilon }_{\alpha }+\sum_n n\Omega W^{(n)}_\alpha,    
\end{aligned}
\label{ }
\end{equation}
where the second line follows from the Schr\" {o}dinger equation, the third from the Floquet ansatz, and the fourth from $\sum_n W^{(n)}_\alpha=1$ . The Brillouin-zone integral of this quantity gives the total mean energy of the Floquet band.

Moreover, it can be shown that the quantum metric of the Floquet states, defined in the time domain, coincides with the \emph{energy fluctuation} in a Floquet band:
\begin{equation}
\begin{aligned}
 &{g_{\alpha }^{tt}}={{ { \langle{E}^{2}\rangle_{\alpha }(t)} }}-   \langle E\rangle_{\alpha } (t)^{2} \\ 
 & ={\langle i{{\partial }_{t}}{{\psi }_{\alpha }}(t)|i{{\partial }_{t}}{{\psi }_{\alpha }}(t)\rangle }  -\langle i{{\partial }_{t}}{{\psi }_{\alpha }}(t)|{{\psi }_{\alpha }}(t)\rangle \langle {{\psi }_{\alpha }}(t)|i{{\partial }_{t}}{{\psi }_{\alpha }}(t)\rangle  \\ 
 & ={\langle i{{\partial }_{t}}{{\phi }_{\alpha }}(t)|i{{\partial }_{t}}{{\phi }_{\alpha }}(t)\rangle } -\langle i{{\partial }_{t}}{{\phi }_{\alpha }}(t)|{{\phi }_{\alpha }}(t)\rangle \langle {{\phi }_{\alpha }}(t)|i{{\partial }_{t}}{{\phi }_{\alpha }}(t)\rangle,
\end{aligned}
\label{gtt}
\end{equation}
where the last equation exploits the gauge independence of the quantum metric. The physical interpretation extends naturally to the other components. The longitudinal component, for example, captures the polarization fluctuation: $g^{xx}_\alpha={{ { \langle{x}^{2}\rangle_{\alpha }(t)} }}-   \langle x\rangle_{\alpha } (t)^{2}$. The mixed component, in turn, encodes the correlation between polarization and energy of the Floquet states:  $g^{xt}_\alpha={{ { \langle x E\rangle_{\alpha }(t)} }}-   \langle x\rangle_{\alpha } (t)  \langle E\rangle_{\alpha }(t)$. Thus, quantum geometry in Floquet systems is central to characterizing a range of physically relevant quantities.

\subsection{Symmetry analysis}\label{symsec}
This subsection investigates the effects of internal and spatial symmetries on the quasienergy spectrum, phase bands, and quantum geometry of Floquet systems. The internal symmetries considered here include time-reversal symmetry (TRS), sublattice symmetry (SS), and particle-hole symmetry (PHS), while the spatial symmetries include parity symmetry (PS), parity-particle-hole symmetry (PPHS), parity-time symmetry (PTS), $n$-fold rotational symmetry, and reflection symmetry.

  \subsubsection{Time-reversal symmtery}\label{TRSsec}
Time reversal inverts both momentum and time. The corresponding operator is anti-unitary and can be written as $\mathcal{T}=u_\mathcal{T} \mathcal{K}$, with $u_\mathcal{T}$ a unitary matrix and  $\mathcal{K}$ the complex conjugation operator. As a result,  $i\partial_\mathbf{k}$ and  $i\partial_t$ are invariant under time reversal. This symmetry is formally defined by 
\begin{equation}
\begin{aligned}
\mathcal{T}H(-\mathbf{k},-t)\mathcal{T}^{-1}=H(\mathbf{k},t).
\end{aligned}
\label{TRScond}
\end{equation}
Equation.~(\ref{TRScond}) implies that if  $\phi (\mathbf{k},t)$ is a Floquet state of $H(\mathbf{k},t)$, then $\mathcal{T}\phi(\mathbf{k},t)$ is a Floquet state of $H(\mathbf{-k},-t)$ sharing the same quasienergy:
\begin{equation}
\begin{aligned}
\text{if}~~~&H(\mathbf{k},t) \phi_\alpha(\mathbf{k},t)=(\epsilon_\alpha(\mathbf{k})+i\partial_t) \phi_\alpha(\mathbf{k},t),\\
\text{then}~~~&H(-\mathbf{k},-t) \mathcal{T}\phi_\alpha(\mathbf{k},t)=(\epsilon_\alpha(\mathbf{k})+i\partial_t) \mathcal{T}\phi_\alpha(\mathbf{k},t).
\end{aligned}
\label{tshart}
\end{equation}
It follows from Eq.~(\ref{tshart}) that TRS enforces symmetry of the quasienergies with respect to $\mathbf{k} \leftrightarrow -\mathbf{k}$. According to the definition of ${\Omega}_\beta^{xy}$ in Eq.~(\ref{omt}) and using Eq.~(\ref{tshart}), for each term like $\langle {{\phi }_{\beta }}(\mathbf{k},t)|{{x}^{e}}|{{\phi }_{\alpha }}(\mathbf{k},t)\rangle $ there exists another term $\langle \mathcal{T}{{\phi }_{\beta }}(\mathbf{k},t)|{{x}^{e}}|\mathcal{T}{{\phi }_{\alpha }}(\mathbf{k},t)\rangle=\langle \mathcal{T}{{\phi }_{\beta }}(\mathbf{k},t)|\mathcal{T}{{x}^{e}}{{\phi }_{\alpha }}(\mathbf{k},t)\rangle=\langle {{\phi }_{\beta }}(\mathbf{k},t)|{{x}^{e}}|{{\phi }_{\alpha }}(\mathbf{k},t)\rangle^*$ where we have used the fact $[\mathcal{T},x^e]=0$ and that for every anti-unitary operator like $\mathcal{T}$ and two arbitrary states we have: $\langle \mathcal{T}\Psi| \mathcal{T} \Phi\rangle =\langle \Psi|  \Phi\rangle^*$. One can write $\mathcal{G}_{+ }^{xy*}(-\mathbf{k},-t)=\mathcal{G}_{+ }^{xy}(\mathbf{k},t)$ and $\mathcal{G}_{+ }^{xt*}(-\mathbf{k},-t)=\mathcal{G}_{+ }^{xt}(\mathbf{k},t)$ where $+$ index stands for upper band. In conclusion, the presence of TRS enforces a vanishing Chern number at $t=0$, and consequently the Chern numbers remain zero at all times, i.e., ${C}^{xy}=0$ and ${C}^{xt}=0$. Furthermore, we have shown that the Brillouin-zone integral of the Fourier components of the Berry curvature vanishes. This result can be extended to the Berry curvature components involving one time index, whose Brillouin-zone integrals also vanish, since the time-reversal operator commutes with $i\partial_t$. Therefore:
\begin{equation}
\begin{aligned}
{C}^{xy}={C}^{xt}=0,~~~&\text{for}~~\text{TRS},\\
\end{aligned}
\label{ }
\end{equation}
and there is also no charge pumping in the presence of TRS.

   \subsubsection{Sublattice symmtery}\label{SSsec}
Under the sublattice (chiral) operation, the sign of time is reversed, while the momentum remains unchanged. The corresponding symmetry condition is formally expressed as
\begin{equation}
\begin{aligned}
u_\mathcal{S}H(\mathbf{k},-t)u_\mathcal{S}^{-1}=-H(\mathbf{k},t),
\end{aligned}
\label{}
\end{equation}
Here $u_\mathcal{S}$  denotes a unitary sublattice operator. The position operator commutes with $[x,u_\mathcal{S}]=0$, whereas $i\partial_t$  anti-commutes,  $\{i\partial_t,u_\mathcal{S}\}=0$. By the same reasoning as in Eq.~(\ref{tshart}), we conclude that for every Floquet state $\phi (\mathbf{k},t)$ of  $H(\mathbf{k},t)$, there exists a partner state  $u_\mathcal{S}\phi(\mathbf{k},t)$ that is a Floquet state of  $H(\mathbf{k},-t)$ with quasienergy of opposite sign:
\begin{equation}
\begin{aligned}
\text{if}~~~&H(\mathbf{k},t) \phi_\alpha(\mathbf{k},t)=(\epsilon_\alpha(\mathbf{k})+i\partial_t) \phi_\alpha(\mathbf{k},t),\\
\text{then}~~~&H(\mathbf{k},-t) u_\mathcal{S}\phi_\alpha(\mathbf{k},t)=(-\epsilon_\alpha(\mathbf{k})+i\partial_t) u_\mathcal{S}\phi_\alpha(\mathbf{k},t).
\end{aligned}
\label{sshart}
\end{equation}
It follows from Eq.~(\ref{sshart}) that SS enforces a pairing of quasienergies at each  $\mathbf{k}$: they come in $\pm$ pairs, symmetric about the momentum axis. Further, as shown in Sec.~\ref{appsym} of the appendix, the phase bands at $t$ and $T-t$ are identical if $U_T=1$ in the presence of SS. 

According to the definition of geometric tensor, we get that for a Hamiltonian with SS $\mathcal{G}_{+ }^{xy}(\mathbf{k},-t)=\mathcal{G}_{- }^{xy}(\mathbf{k},t)$, $\mathcal{G}_{+ }^{xt}(\mathbf{k},-t)=-\mathcal{G}_{- }^{xt}(\mathbf{k},t)$ where the subscript $+(-)$ stands for upper (lower) band. Since the total Chern number summed over all bands vanishes, each Floquet band must have zero Chern number. Consequently, we have
\begin{equation}
\begin{aligned}
 C^{xy}=0,~~~&\text{for}~~\text{SS}.
\end{aligned}
\label{ }
\end{equation}

 \subsubsection{Particle-hole symmetry}\label{PHSsec}
PHS, also known as charge conjugation, inverts the momentum but preserves time. The formal definition is
\begin{equation}
\begin{aligned}
\mathcal{C}H(-\mathbf{k},t)\mathcal{C}^{-1}=-H(\mathbf{k},t),
\end{aligned}
\label{phscond}
\end{equation}
where $\mathcal{C}$ denotes an anti-unitary particle-hole operation. Accordingly, the position operator satisfies $[x,\mathcal{C}]=0$, while $i\partial_t$  obeys the anti-commutation relation  $\{i\partial_t,\mathcal{C}\}=0$. As a result,
\begin{equation}
\begin{aligned}
\text{if}~~~&H(\mathbf{k},t) \phi_\alpha(\mathbf{k},t)=(\epsilon_\alpha(\mathbf{k})+i\partial_t) \phi_\alpha(\mathbf{k},t),\\
\text{then}~~~&H(-\mathbf{k},t) \mathcal{C}\phi_\alpha(\mathbf{k},t)=(-\epsilon_\alpha(\mathbf{k})+i\partial_t) \mathcal{C}\phi_\alpha(\mathbf{k},t).
\end{aligned}
\label{phshart}
\end{equation}
Equation~(\ref{phshart}) shows that PHS enforces a pairing of quasienergies: for each quasienergy at momentum $\mathbf{k}$, there exists an equal but opposite quasienergy at $-\mathbf{k}$.

As shown in Sec.~\ref{appsym} of the appendix, the PHS ensures that for each phase band at momentum $\mathbf{k}$, there exists a corresponding phase band with the opposite value at $-\mathbf{k}$. These PHS-conjugate pairs carry Chern numbers of equal magnitude and opposite sign. The PHS implies $\mathcal{G}_{+ }^{xy*}(-\mathbf{k},t)=\mathcal{G}_{- }^{xy}(\mathbf{k},t)$, $\mathcal{G}_{+ }^{xt*}(-\mathbf{k},t)=-\mathcal{G}_{- }^{xt}(\mathbf{k},t)$.  Consequently
\begin{equation}
\begin{aligned}
C^{xt}=0,~~~&\text{for}~~\text{PHS}.
\end{aligned}
\label{ }
\end{equation}

\begin{table*}[]
\caption{Symmetry constraints on quantum geometry of Floquet states}
\begin{tabular}{lll}
\hline
symmetry &   constraint on quantum geometry & consequence  \\ \hline
TRS   & $\mathcal{G}_{+ }^{xy*}(-\mathbf{k},-t)=\mathcal{G}_{+ }^{xy}(\mathbf{k},t)$, $\mathcal{G}_{+ }^{xt*}(-\mathbf{k},-t)=\mathcal{G}_{+ }^{xt}(\mathbf{k},t)$ & $C^{xy}=C^{xt}=0$            \\
SS             & $\mathcal{G}_{+ }^{xy}(\mathbf{k},-t)=\mathcal{G}_{- }^{xy}(\mathbf{k},t)$, $\mathcal{G}_{+ }^{xt}(\mathbf{k},-t)=-\mathcal{G}_{- }^{xt}(\mathbf{k},t)$ & $C^{xy}=0 $              \\
PHS       & $\mathcal{G}_{+ }^{xy*}(-\mathbf{k},t)=\mathcal{G}_{- }^{xy}(\mathbf{k},t)$, $\mathcal{G}_{+ }^{xt*}(-\mathbf{k},t)=-\mathcal{G}_{- }^{xt}(\mathbf{k},t)$ & $C^{xt}=0$            \\                 
PS    &$\mathcal{G}_{+ }^{xy}(-\mathbf{k},t)=\mathcal{G}_{+ }^{xy}(\mathbf{k},t)$, $\mathcal{G}_{+ }^{xt}(-\mathbf{k},t)=-\mathcal{G}_{+ }^{xt}(\mathbf{k},t)$  &   $q^{xt}(t)=C^{xt}=0$              \\
PTS             &$\mathcal{G}_{+ }^{xy*}(\mathbf{k},-t)=\mathcal{G}_{+ }^{xy}(\mathbf{k},t)$, $\mathcal{G}_{+ }^{xt*}(\mathbf{k},-t)=-\mathcal{G}_{+ }^{xt}(\mathbf{k},t)$  &  $ \Omega^{xy}(t)+\Omega^{xy}(-t)=0, g^{xt}(t)+g^{xt}(-t)=0$                 \\
$C_{n\ge3}$      &    Eq.~(\ref{GC3})   & $q^{xx}(t)=q^{yy}(t), q^{xy}(t)=q^{xt}(t)=C^{xt}=0$                 \\
$\mathcal{R}_x$ &  $\mathcal{G}_{+ }^{xy}(-k_x,t)=-\mathcal{G}_{+}^{xy}(\mathbf{k},t)$, $\mathcal{G}_{+ }^{xt}(-k_x,t)=-\mathcal{G}_{+ }^{xt}(\mathbf{k},t)$ & $q^{xy}(t)=q^{xt}(t)=0$,  $C^{xy}=C^{xt}=0$ \\
\end{tabular}
\label{symtab}
\end{table*}

 \subsubsection{Parity, parity-time and parity-particle-hole symmetry}\label{PTSsec}
PS is a key unitary symmetry implemented by the unitary operator $u_\mathcal{P}$. It inverts the momentum while preserving the sign of time. The formal symmetry condition is
\begin{equation}
\begin{aligned}
u_\mathcal{P}H(\mathbf{k},t)u_\mathcal{P}^{-1}=H(\mathbf{-k},t).
\end{aligned}
\label{PScond}
\end{equation}
Under parity, the position operator ($i\partial_\mathbf{k}$) satisfies the anti-commutation relation $\{x,u_\mathcal{P}\}=0$, whereas $i\partial_t$ commutes, $[i\partial_t,u_\mathcal{P}]=0$. The consequences of this symmetry are
\begin{equation}
\begin{aligned}
\text{if}~~~&H(\mathbf{k},t) \phi_\alpha(\mathbf{k},t)=(\epsilon_\alpha(\mathbf{k})+i\partial_t) \phi_\alpha(\mathbf{k},t),\\
\text{then}~~~&H(-\mathbf{k},t) u_\mathcal{P}\phi_\alpha(\mathbf{k},t)=(\epsilon_\alpha(\mathbf{k})+i\partial_t) u_\mathcal{P}\phi_\alpha(\mathbf{k},t).
\end{aligned}
\label{pshart}
\end{equation}
Hence, the PS implies that the quasienergies are invariant under $\mathbf{k} \leftrightarrow \mathbf{-k}$, with the corresponding wavefunctions connected by the operator $u_\mathcal{P}$.

Furthermore, in a parity-symmetric Hamiltonian, the phase bands are symmetric with respect to $\mathbf{k} \leftrightarrow \mathbf{-k}$ as proved in appendix.
It is evident that in the presence of PS:  $\mathcal{G}_{+ }^{xy}(-\mathbf{k},t)=\mathcal{G}_{+ }^{xy}(\mathbf{k},t)$, $\mathcal{G}_{+ }^{xt}(-\mathbf{k},t)=-\mathcal{G}_{+ }^{xt}(\mathbf{k},t)$.   As a consequence
\begin{equation}
\begin{aligned}
q^{xt}(t)=C^{xt}=0,~~~&\text{for}~~\text{PS}.\\
\end{aligned}
\label{qxtzero}
\end{equation}
Thus, when the PS is present, the correlation between the energy and polarization of the Floquet states, integrated over the Brillouin zone, is identically zero. Charge pumping is therefore forbidden. It is noteworthy that a basic spatial symmetry can nullify the topological invariant $C^{xt}$, an aspect that has received relatively little attention in earlier studies.

Parity-time operation preserves momentum but inverts time. Its formal definition reads
\begin{equation}
\begin{aligned}
\mathcal{PT}H(\mathbf{k},-t)\mathcal{PT}^{-1}=H(\mathbf{k},t).
\end{aligned}
\label{}
\end{equation}
where  $\mathcal{PT}$ denotes an anti-unitary operator that satisfies $\{x,\mathcal{PT}\}=0$  and $[i\partial_t,\mathcal{PT}]=0$. Repeating the analysis of the preceding sections, we obtain for a PT-symmetric model:
\begin{equation}
\begin{aligned}
\text{if}~~~&H(\mathbf{k},t) \phi_\alpha(\mathbf{k},t)=(\epsilon_\alpha(\mathbf{k})+i\partial_t) \phi_\alpha(\mathbf{k},t),\\
\text{then}~~~&H(\mathbf{k},-t) \mathcal{PT}\phi_\alpha(\mathbf{k},t)=(\epsilon_\alpha(\mathbf{k})+i\partial_t) \mathcal{PT}\phi_\alpha(\mathbf{k},t).
\end{aligned}
\label{ptshart}
\end{equation}
Hence, in a PT-symmetric system, $\mathcal{G}_{+ }^{xy*}(\mathbf{k},-t)=\mathcal{G}_{+ }^{xy}(\mathbf{k},t)$ and $\mathcal{G}_{+ }^{xt*}(\mathbf{k},-t)=-\mathcal{G}_{+ }^{xt}(\mathbf{k},t)$. This forces the Berry curvatures at $t$ and $-t$ to cancel, $\Omega_\beta^{xz}(t)+\Omega_\beta^{xz}(-t)=0$, so that the Chern number vanishes: $C_\beta=0$. Additionally, the Berry curvature is zero at $t=0, T$. Since the parity-time operator commutes with $i\partial_t$, it follows that
\begin{equation}
\begin{aligned}
 g^{xt}(t=0,T)=g^{yt}(t=0,T)=0,~~~&\text{for}~~\text{PTS},\\
 \Omega^{xy}(t=0,T;\mathbf{k})=0,~~~&\text{for}~~\text{PTS}.
\end{aligned}
\label{}
\end{equation}

Another symmetry, PPHS, can be defined with an antiunitary operator  $\mathcal{PP}$ via the condition
\begin{equation}
\begin{aligned}
\mathcal{PP}H(\mathbf{k},t)\mathcal{PP}^{-1}=-H(\mathbf{k},t).
\end{aligned}
\label{pphseq}
\end{equation}
This condition enforces symmetry of both the Floquet quasienergies and the phase bands about the momentum axis at all times. As a result, the bands appear in pairs with opposite quasienergies, opposite Berry curvatures, and equal quantum metrics.

\subsubsection{$n$-fold rotational and reflection symmetry}
Consider a Hamiltonian with $n$-fold rotational symmetry, where  $n \ge 3$. For simplicity, we specialize to $n=3$. The condition defines this symmetry
\begin{equation}
\begin{aligned}
 \mathcal{C}_3 H(\mathbf{k},t)\mathcal{C}_3^{-1}=H(\mathbf{k}^\prime,t),
\end{aligned}
\label{cncond}
\end{equation}
 with $\mathcal{C}_3$ a unitary operator and $(k_x^\prime,k_y^\prime)=\frac{1}{2}(-k_x-\sqrt{3}k_y, \sqrt{3}k_x-k_y)$.  Thus
\begin{equation}
\begin{aligned}
\text{if}~~~&H(\mathbf{k},t) \phi_\alpha(\mathbf{k},t)=(\epsilon_\alpha(\mathbf{k})+i\partial_t) \phi_\alpha(\mathbf{k},t),\\
\text{then}~~~&H(\mathbf{k}^\prime,t) \mathcal{C}_3 \phi_\alpha(\mathbf{k},t)=(\epsilon_\alpha(\mathbf{k})+i\partial_t) \mathcal{C}_3 \phi_\alpha(\mathbf{k},t).
\end{aligned}
\label{cnshart}
\end{equation}

Therefore, the quasienergies are symmetric with respect to $n$-fold rotation in $k$ space. We show in appendix that for a system possessing at least threefold rotational symmetry, we have
\begin{equation}
\begin{aligned}
q^{xx}(t)=q^{yy}(t),~~~&\text{for}~~C_{n \ge 3}~~\text{symmetric},\\
q^{xy}(t)=q^{xt}(t)=C^{xt}=C^{yt}=0,~~~&\text{for}~~C_{n \ge 3}~~\text{symmetric}.
\end{aligned}
\label{cnimp}
\end{equation}

An analogous reasoning holds for reflection symmetry—say, about $x=0$—defined by a unitary operator $\mathcal{R}_x$  through
\begin{equation}
\begin{aligned}
 \mathcal{R}_x H(k_x,k_y,t)\mathcal{R}_x^{-1}=H(-k_x,k_y,t).
\end{aligned}
\label{}
\end{equation}
Under this operation, the $x^e$  coordinate acquires a minus sign, while  $y^e$ and $i\partial_t$  remain unaffected, which yields
\begin{equation}
\begin{aligned}
q^{xy}(t)=q^{xt}(t)=C^{xy}=C^{xt}=0,~~~&\text{for}~~\mathcal{R}_x~\text{symmetric}.\\
\end{aligned}
\label{rximp}
\end{equation}
Table~\ref{symtab} summarizes the constraints placed on the quantum geometric tensor of Floquet systems by different symmetries, along with their implications. It is hoped that this summary will prove useful for the design and engineering of quantum geometry in Floquet systems.

\section{Numerical results}\label{num}
In this section, the Berry curvature, Chern number, and the Brillouin-zone-integrated quantum metric for the RLBL model \cite{rudner2013anomalous}, a simple yet illustrative example, are calculated. The model is formulated on a square lattice of unit spacing, with two spinless atoms per unit cell. With the vectors  $b_1=-b_3=(1,1)$ and $b_2=-b_4=(-1,1)$, the momentum-space Hamiltonian takes the form
\begin{equation}
\begin{aligned}
&{{H}^{RLBL}}(\mathbf{k},t)=h_i,~~~\text{for}~~ (i-1)T/5<t<(i)T/5,\\
&h_i={J_i({{e}^{i\mathbf{k}.{{\text{b}}_{i}}}}{{\sigma }_{+}}+{{e}^{-i\mathbf{k}.{{\text{b}}_{i}}}}{{\sigma }_{-}})+\delta {{\sigma }_{z}}}
\end{aligned}
\label{hrlbl}
\end{equation}
where $J_i$  is the hopping amplitude, $\delta$  the sublattice potential and $\sigma_{\pm}=\sigma_x \pm i \sigma_y$. The time evolution consists of five steps: for $(i-1)T/5<t<(i)T/5$, ~$i=1,2,3,4$ we have $J_i=J$ and a nonzero $\delta$, while in the last step, $4T/5<t<T$, only the sublattice potential acts ($J_5=0$). The Brillouin zone of the RLBL model (Eq.~(\ref{hrlbl})) is a square with $-\pi<k_{x,y}<\pi$.

When $\delta \neq 0$, the model lacks any internal symmetry and falls into class $A$ of the Floquet topological insulator periodic table \cite{harper2020topology}, where the topology is characterized by the gap winding number (\ref{w3eq}) and the Floquet-band Chern number. It does, however, respect PPHS as per Eq.~(\ref{pphseq}) with $\mathcal{PP}=\sigma_y \mathcal{K}$.

Only at $\delta=0$ the model additionally exhibits PHS (Eq.~(\ref{phscond}) with $\mathcal{C}=\sigma_z\mathcal{K}$ and PS (Eq.~(\ref{pshart}) with $u_\mathcal{P}=\sigma_x$).

\begin{figure}
\includegraphics[width=\linewidth,trim={0 0 0.25cm 0}, clip]{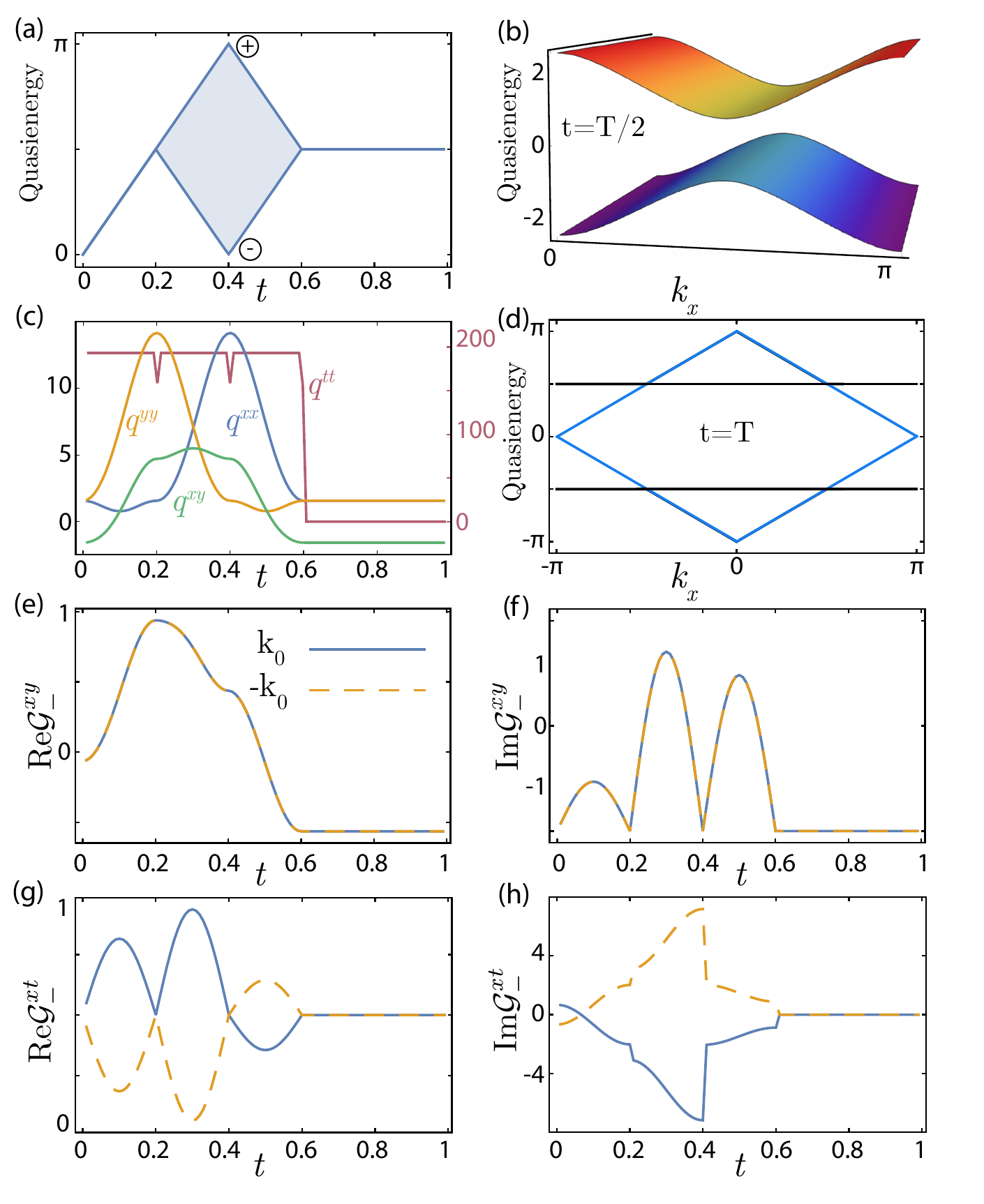} 
\caption{(a) Time-dependent minimum and maximum of the highest phase band (shaded region) for the RLBL model. Gap closings occur at $t=0.4T$ which is a topological singularity. The topological charge of each gap closing is shown in a nearby circle. (b) The phase bands at $t=T/2.$ (c) Brillouin-zone-integrated quantum metric components  ${q}^{mn}=\sum_\mathbf{k}{g}^{mn}$ where $\{m,n\}\in \{x,y,t\} $ versus time. The PS-enforced zeros $q^{xt}(t)=q^{yt}(t)=C^{xt}=C^{yt}=0$  are not displayed. (d) Nanoribbon band structure of the RLBL model at $t=T$, with edge states highlighted in blue. (e-h) The quantum geometric tensors $\mathcal{G}_-^{xy}$ and $\mathcal{G}_-^{xt}$ at two special momenta $\mathbf{k}=-\mathbf{k}_0$ and $ \mathbf{k}= \mathbf{k}_0=(0.1,0.2)$ at given $J=5\pi/2, T=1, \delta=0$.}
\label{rlbl}
\end{figure}

The evolution operator over one period takes the form
\begin{equation}
\begin{aligned}
U_T^{RLBL}=e^{-ih_5T/5}e^{-ih_4T/5}e^{-ih_3T/5}e^{-ih_2T/5}e^{-ih_1T/5},
\end{aligned}
\label{}
\end{equation}
from which the Floquet states and quasienergies are obtained by using diagonalization method, according to Eq.~(\ref{UTeq}).

\begin{figure}
\includegraphics[width=\linewidth,trim={0 0 0.25cm 0}, clip]{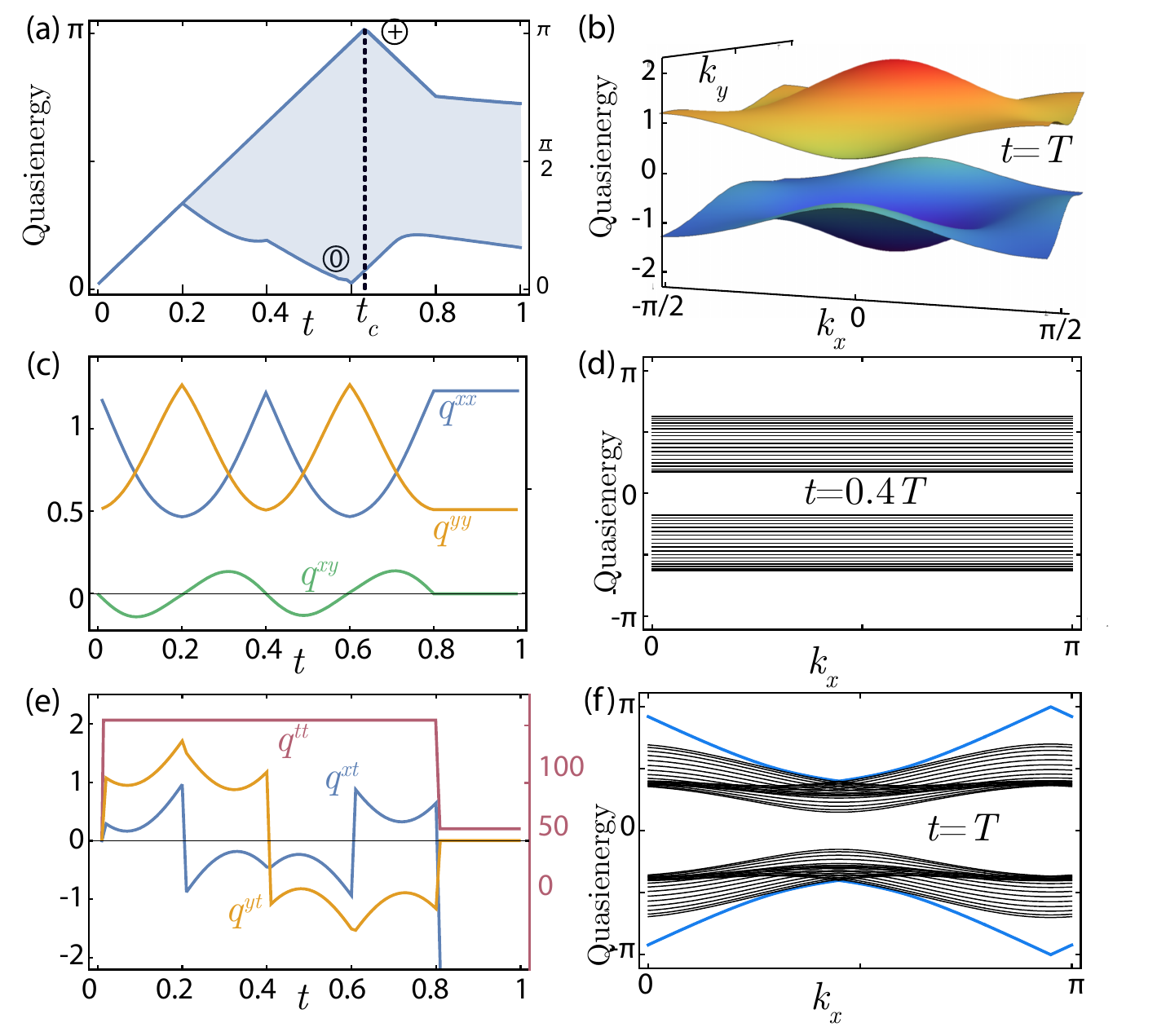} 
\caption{(a) Time evolution of the highest phase band extrema (shaded envelope) for the RLBL model. Each gap closing is labeled with its topological charge in a circle; the $+1$ charge at $t=t_c$  signals the appearance of a dynamical edge state. (b) Bulk Floquet quasienergy spectrum of the RLBL model. (c),(e) Time dependence of the integrated quantum metric components ${q}^{mn}=2\pi\sum_\mathbf{k}{g}^{mn}$ where $\{m,n\}\in \{x,y,t\}$, with the scale for  $q^{tt}$ shown on the right axis.  (d),(f) Nanoribbon dispersion at $t=0.4T,T$; edge states appear in blue in (f). Parameters: $J=3\pi/2, T=1, \delta=0.5\pi/T$.}
\label{rlbl2}
\end{figure}

 The maximum and minimum of the topmost phase band of the RLBL model are shown in Fig.~\ref{rlbl}(a) for the parameter $J_1=J_2=J_3=5\pi/2, J_4=J_5=0, \delta=0$, and $T=1$, obtained by diagonalizing the evolution operator according to Eq.~(\ref{utdef}). The lower band is not displayed because the RLBL model possesses PPHS, which enforces symmetry of the phase bands and quasienergies about the momentum axis—a feature also evident in Figs.~\ref{rlbl}(b),(d) and \ref{rlbl2}(b),(f). Moreover, at $\delta=0$ the model additionally exhibits PS, making the quasienergies and phase bands symmetric under $\mathbf{k}\leftrightarrow -\mathbf{k}$. The  phase bands at the mid-cycle $(t=T/2)$ are depicted in Fig.~\ref{rlbl}(b) respecting the aforementioned symmetries. For these parameters, the quasienergy bands are flat and independent of $\mathbf{k}$. This is also visible in Fig.~\ref{rlbl}(d), which depicts the nanoribbon dispersion: bulk quasienergies being flat horizontal lines and two edge states (shown in blue) crossing the gap. Notably, Fig.~\ref{rlbl}(a) reveals that the phase bands are completely flat for $0<t<T/5$ and $3T/5<t<T$. The gap closes at $t=2T/5$ marking a topological phase transition. At this point, the gaps at quasienergies $0$ and $\pi$ close simultaneously, carrying opposite topological charges of $+1$ and $-1$ as indicated by the circled numbers. Consequently, edge modes are absent (present) for $t<2T/5 (t>2T/5)$.

The Brillouin-zone-integrated quantum metric, or \emph{quantum weight}, evaluated in the Floquet-state basis: ${q}^{mn}=2
\pi\sum_{\mathbf{k}}{g}^{mn}$, $\{m,n\}\in \{x,y\}$ is shown in Fig.~\ref{rlbl}(b). The quantity is free of divergences and traces out a smooth oscillatory curve that becomes flat for $t>0.6T$. In particular, the component $q^{xx} (q^{yy})$ reaches its peak at $t=0.4T (0.2T)$. Consistent with expectation, the Floquet-basis quantum weight is single-valued at $t=0$ and $t=T$.

\begin{figure*}
\includegraphics[width=\linewidth,trim={0 0 0.25cm 0}, clip]{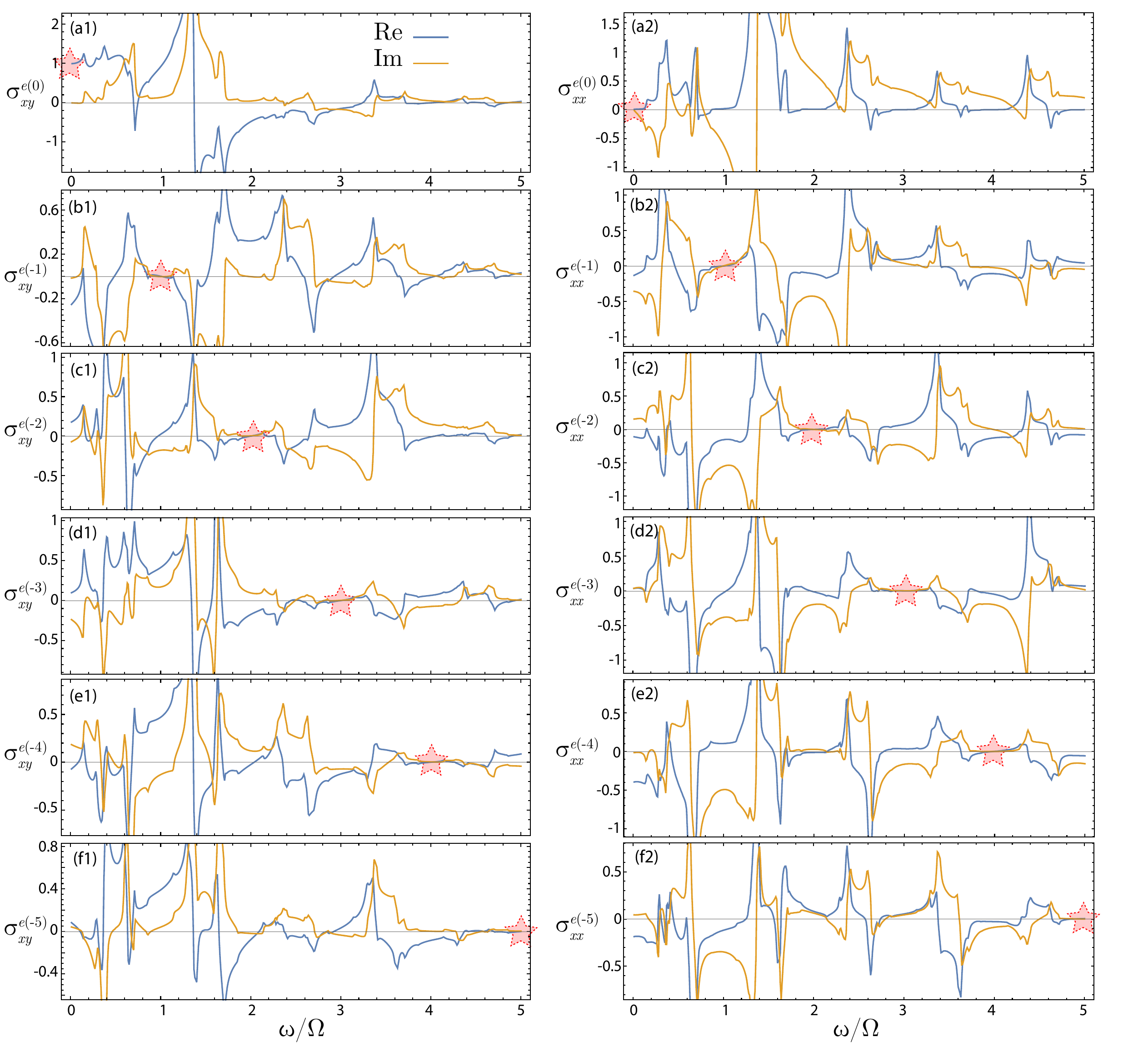} 
\caption{(a)–(f) Frequency dependence of the Fourier components $\sigma^{[1]e(-n)}_{xy}$  and $\sigma^{[1]e(-n)}_{xx},~~n=0,1,2,3,4,5$ of the optical conductivity for the RLBL model, shown in units of $e^2/h$ under the assumption of ideal Floquet-state occupation. The quantized DC value—proportional to the Chern number—and the conductivity zeros mandated by Eqs.~(\ref{xzzero}) and (\ref{xxzero}) are marked by dashed red stars. The parameters are $J=3\pi/2, T=1, \delta=0.5\pi/T$.}
\label{ocfig}
\end{figure*}

For the sake of completeness, Fig.~\ref{rlbl}(c) also displays (in red) the time-domain quantum weight $q^{tt}$, defined by integrating the quantum metric of Eq.~(\ref{gtt}) over the Brillouin zone. This quantity remains free of divergences, as expected for a basis related by a smooth unitary transformation. The components $q^{xt}, q^{yt}$ are identically zero by PS (Eq.~(\ref{qxtzero})) and are omitted, reflecting the vanishing correlation between polarization and energy of the Floquet states. Similarly, PS enforces $C^{xt}=C^{yt}=0$, so there is no charge pumping.

For the RLBL model with $\delta=0$, we numerically verify the symmetry-imposed constraints on the geometric tensor that stem 
from PS and PHS, as catalogued in Table~\ref{symtab}. The corresponding numerical results are depicted in Figs.~\ref{rlbl}(e–h). As shown in Figs.~\ref{rlbl}(e,f), the real and imaginary components of $\mathcal{G}_-^{xy}$ coincide for pairs of the opposite momenta i.e. $\mathcal{G}_-^{xy}(\mathbf{k}_0)=\mathcal{G}_-^{xy}(-\mathbf{k}_0)$, where $\mathbf{k}_0=(0.1,0.2)$. In addition, the geometric tensor component satisfies $\mathcal{G}_-^{xt}(\mathbf{k}_0)=-\mathcal{G}_-^{xt}(-\mathbf{k}_0)$, as illustrated in Figs.~\ref{rlbl}(g,h); this result is in full accord with the PS constraints listed in the second column of Table~\ref{symtab}. Given that the model under consideration possesses two bands, the relation $\mathcal{G}_+=\mathcal{G}_-^*$   holds, which independently reaffirms the PHS constraints presented in Table~\ref{symtab}. Lastly, the quantum weights $q^{xt}(t), q^{yt}(t)$, and the mixed Chern numbers $C^{xt}, C^{yt}$  are all found to vanish, consistently with the predictions of the third column of Table~\ref{symtab}.

Next, we consider a different set of parameters for the RLBL model that gives rise to a nontrivial one-cycle evolution and a finite Floquet-band Chern number. The extrema of the uppermost phase band as a function of time are shown in Fig.~\ref{rlbl2}(a) for the parameter set $J=3\pi/2$, $T=1$, and $\delta=0.5\pi/T$. Remarkably, the resulting Floquet band acquires a Chern number of unity, even though the instantaneous Hamiltonian preserves TRS at every instant during the evolution. The bulk quasienergy spectrum is presented in Fig.~\ref{rlbl2}(b), while the corresponding nanoribbon spectra are shown in Figs.~\ref{rlbl2}(d) and \ref{rlbl2}(f) at $t=0.4T$ and $t=T$, respectively. These spectra confirm the emergence of chiral edge states crossing the dynamical gap for $t>t_c$. The zone-edge gap closing at $t=t_c$ carries a topological charge of $+1$ (indicated by the circled number), giving rise to the nonzero Floquet-band Chern number. In contrast, the gap closing at zero quasienergy that occurs before $t_c$ is topologically trivial, carrying zero topological charge. Furthermore, for $\delta \neq 0$, PS is broken, and consequently the spectrum in Fig.~\ref{rlbl2}(f) is no longer symmetric under the transformation $\mathbf{k}\leftrightarrow -\mathbf{k}$. The edge modes are highlighted in blue.

The quantum weight $q^{mn}$, evaluated in the Floquet-state basis, is presented in Figs.~\ref{rlbl2}(c) and (e). The spatial components of the quantum weight evolve continuously and exhibit an oscillatory time dependence, whereas the components involving the temporal index display discontinuities at specific instants. The mixed space-time Chern numbers satisfy $C^{xt}=C^{yt}=0$ and are therefore omitted from the figure. In general, the quantum metric and Berry curvature of an isolated band obey the inequalities
\begin{eqnarray}
\mathrm{Tr}(q_\beta)=q^{xx}_\beta+q^{yy}_\beta &\ge& C_\beta,
\label{eq:trace_bound}\\
\sqrt{\det(q_\beta)}
=\sqrt{q^{xx}_\beta q^{yy}_\beta-(q^{xy}_\beta)^2}
&\ge& \frac{C_\beta}{2},
\label{eq:det_bound}
\end{eqnarray}
which are manifestly satisfied throughout the evolution. Since the Floquet band retains a Chern number of unity at all times, the trace of the quantum weight is bounded from below by one, in agreement with Fig.~\ref{rlbl2}(c). The temporal component $q^{tt}$ is shown in dark red in Fig.~\ref{rlbl2}(e), with its magnitude indicated on the right-hand axis.

We would like to emphasize a direct experimental path for reconstructing the time-dependent quantum geometry of Floquet-Bloch states with the relations discussed here. In a pump-probe experiment, the Floquet state is initially prepared by a strong periodic drive, and the frequency- and time-resolved optical conductivity is measured by a mild probing field. The various harmonic components $\sigma^{(n)}(\omega)$ can be separated by Fourier decomposing the observed conductivity with respect to the drive period. The relevant Fourier components of the quantum geometric tensor can then be reconstructed from the recorded optical spectra using the sum rules of Eqs. (27) and (28). Additionally, because of their different dependency on the probe polarization, the symmetric (quantum metric) and antisymmetric (Berry curvature) contributions can be accessed separately using polarization-resolved spectroscopy, especially circular dichroism. Thus, utilizing current ultrafast pump-probe, terahertz conductivity, and angle-resolved photoemission (tr-ARPES) techniques, our formalism offers both a theoretical description of Floquet quantum geometry and a useful framework for physically measuring its time development.

\subsection{Optical conductivity}\label{}
To complete our discussion, we calculate the interband optical conductivity of the RLBL model using Eq.~(\ref{oc1}). We assume an ideal occupation of the Floquet states, in which the lower (upper) Floquet band within the FBZ is completely filled (empty). Under this assumption, the intraband contribution to the optical conductivity vanishes~\cite{dabiri2025dynamical}. Fig.~\ref{ocfig} presents several Fourier components of the optical conductivity, $\sigma^{[1]e(-n)}_{xy}$ and $\sigma^{[1]e(-n)}_{xx}$ with $n=0,1,2,3,4,5$, for the RLBL model using the same parameters as in Fig.~\ref{rlbl2}. As shown in Fig.~\ref{ocfig}(a1), the real part of the dc Hall conductivity is quantized and proportional to the Floquet-band Chern number, $C_2=1$, as indicated by the dashed red star. The higher-order Fourier components, displayed in Figs.~\ref{ocfig}(b--f), each vanish at their corresponding harmonic frequencies, in agreement with Eqs.~(\ref{xzzero}) and~(\ref{xxzero}), as highlighted by the dashed red stars.

The time-averaged longitudinal optical conductivity is shown in Fig.~\ref{ocfig}(a2). Notably, the real part becomes negative over a finite frequency range, a distinctive feature of Floquet systems that has no counterpart in static equilibrium systems. This behavior originates from the ideal occupation of the Floquet states, which allows downward optical transitions between Floquet bands~\cite{dabiri3}. A negative real part of the longitudinal conductivity indicates that the probe field is amplified as it propagates through the system, rather than being dissipated. Furthermore, the optical conductivity remains finite even at probe frequencies as high as $\omega \approx 5\Omega$, reflecting the finite spectral weight of the Floquet sidebands and the persistence of photon-assisted optical transitions at high harmonics~\cite{dabiri3}.

As seen from Fig.~\ref{ocfig}, the higher Fourier components of the optical conductivity are of the same order of magnitude as the time-averaged component. This can be attributed to the quench-like nature of the drive, which itself contains multiple Fourier components. For a detailed discussion of photon-assisted optical transitions between Floquet bands, the Floquet weight of the side bands, and their connection to the optical conductivity—including the positions of peaks and dips—we refer the reader to \cite{dabiri3}.

Before we wrap up our results, we would like to emphasize that the main finding of this work is that the various physical phenomena seen in periodically driven systems are unified by a single geometric object: the time-dependent quantum geometric tensor of Floquet-Bloch states.
Importantly, this work shows that a single geometric object, the time-dependent quantum geometric tensor of Floquet-Bloch states, unifies the different physical phenomena that arise in periodically driven systems. While its symmetric component, the quantum metric, describes the local distance between nearby Floquet states and establishes quantities like optical transition strengths, energy fluctuations, and polarization-energy correlations, its antisymmetric component, the Berry curvature, controls anomalous transport, Hall responses, topological invariants, and quantized charge pumping. While the mixed momentum-time components of this tensor naturally encode nonadiabatic charge transport, the optical conductivity sum rules derived in this study show that linear optical responses are directly controlled by the Fourier components of this tensor. Similarly, it is possible to think of the symmetry constraints found for various internal and crystallographic symmetries as limitations placed on the same underlying geometric tensor. This unified viewpoint reveals that optical conductivity, quantum fluctuations, topological pumping, and symmetry-protected responses are all physical manifestations of a single time-dependent geometric structure defined on the manifold of Floquet-Bloch states.

\section{Conclusion}\label{conclusion}
We have developed a unified framework for the quantum metric and time-dependent Berry curvature of Floquet-Bloch states. The main finding of this work is that the various physical phenomena that arise in periodically driven systems, such as optical conductivity, quantum fluctuations, topological pumping, and symmetry-protected responses, can be captured by a single geometric object, the time-dependent quantum geometric tensor. Our approach naturally reduces to static Bloch-band geometry in the zero-frequency limit, preserves the full micromotion, and establishes a direct connection to time-resolved experimental observables by defining this tensor in the Floquet-state basis.

Several important conclusions are drawn from our analysis. We obtained optical sum rules [Eqs. (\ref{resms}) and (\ref{imsps})] that relate the optical conductivity to the Fourier components of the quantum geometric tensor. The first-order DC Hall and longitudinal responses at harmonic frequencies vanish identically under ideal Floquet-band occupations. This conclusion is statistically confirmed in the RLBL model up to the fifth harmonic. We demonstrated that quantized charge pumping happens intrinsically throughout each driving cycle by introducing a mixed Berry curvature. This generalizes Thouless pumping to the non-adiabatic domain without the need for additional biases or precisely calibrated protocols. We also demonstrated how the quantum metric provides a geometric interpretation of quantum fluctuations in Floquet systems by encoding energy fluctuations, polarization fluctuations, and polarization-energy correlations. 

Lastly, our thorough symmetry analysis, which is summarized in Table~\ref{symtab}, shows how the quantum geometric tensor is constrained by time-reversal, sublattice, particle-hole, parity, rotational, and reflection symmetries. This has direct implications for topological invariants and optical selection criteria.

 The analytical predictions are validated by simulations of a fully symmetric Floquet model and the RLBL model. The expected conductivity zeros and the optical sum rules are confirmed by the calculated time-dependent quantum geometric quantities. The ideal occupation assumption and the topological nature of the driven phases are consistent with the appearance of chiral edge states and the detection of negative longitudinal conductivity, a characteristic of Floquet systems.

 Here, time-dependent quantum geometry is established as a flexible Floquet matter diagnostic tool. The sum rules obtained here provide a useful method for reconstructing the quantum geometric tensor from terahertz conductivity measurements and polarization-resolved pump-probe spectroscopy. In ultracold atomic gases, photonic waveguide arrays, or driven two-dimensional materials, where Floquet-state preparation and detection are becoming more and more possible, the non-adiabatic charge pumping method may be implemented. This geometric framework and its applications to quantum materials and devices are expected to be significantly enhanced by future extensions to interacting systems, disordered environments, and non-Hermitian Floquet situations.

\newpage
\appendix

\section{Gauge invariance of quantum metric in Floquet basis}
This section establishes the gauge invariance of the quantum metric introduced in Eq.~(\ref{gt}). To do so, we multiply each Floquet state by a phase factor: $|\phi_\alpha(t)\rangle \rightarrow e^{i\theta_\alpha}|\phi_\alpha(t)\rangle$ and $|\phi_\beta(t)\rangle \rightarrow e^{i\theta_\beta}|\phi_\beta(t)\rangle$. The following term for $\alpha\neq\beta$ is changed as
\begin{equation}
\begin{aligned}
  & \langle {{\phi }_{\beta }}(t)|{{e}^{-i{{\theta }_{\beta }}}}{\partial_\mathbf{k}}{{e}^{i{{\theta }_{\alpha }}}}|{{\phi }_{\alpha }}(t)\rangle  \\ 
 & =i\langle {{\phi }_{\beta }}(t)|{{e}^{-i{{\theta }_{\beta }}}}\{{{\partial }_{\mathbf{k}}}({{e}^{i{{\theta }_{\alpha }}}})|{{\phi }_{\alpha }}(t)\rangle +{{e}^{i{{\theta }_{\alpha }}}}{{\partial }_{\mathbf{k}}}|{{\phi }_{\alpha }}(t)\rangle \} \\ 
 & =i\langle {{\phi }_{\beta }}(t)|{{e}^{-i{{\theta }_{\beta }}}}\{{{e}^{i{{\theta }_{\alpha }}}}{{\partial }_{\mathbf{k}}}|{{\phi }_{\alpha }}(t)\rangle \} \\ 
 & ={{e}^{i({{\theta }_{\alpha }}-{{\theta }_{\beta }})}}\langle {{\phi }_{\beta }}(t)|{\partial_\mathbf{k}}|{{\phi }_{\alpha }}(t)\rangle.
\end{aligned}
\label{nesfberry}
\end{equation}
To obtain the third line we exploited the orthonormality of the Floquet states at each momentum and at every instant of time. With the help of Eq.~(\ref{nesfberry}), one readily checks that the quantum metric of Eq.~(\ref{gt}) is invariant under the multiplication of each Bloch wave function by an arbitrary phase factor, and is therefore gauge invariant.

Let us also verify the gauge invariance of the time-domain quantum metric  $q^{tt}$  introduced in Eq.~(\ref{gtt}). By inserting a complete set of states, Eq.~(\ref{gtt}) can be recast as
\begin{equation}
\begin{aligned}
  & {{q}^{tt}_{\alpha }}=\{\langle {{\partial }_{t}}{{\phi }_{\alpha }}(t)|{{\partial }_{t}}{{\phi }_{\alpha }}(t)\rangle  \\ 
 & -\langle {{\partial }_{t}}{{\phi }_{\alpha }}(t)|{{\phi }_{\alpha }}(t)\rangle \langle {{\phi }_{\alpha }}(t)|{{\partial }_{t}}{{\phi }_{\alpha }}(t)\rangle \} \\ 
 & =\sum\limits_{\beta \ne \alpha }{\langle {{\partial }_{t}}{{\phi }_{\alpha }}(t)|{{\phi }_{\beta }}(t)\rangle \langle {{\phi }_{\beta }}(t)|{{\partial }_{t}}{{\phi }_{\alpha }}(t)\rangle }. \\ 
\end{aligned}
\label{}
\end{equation}
Clearly, in this representation an arbitrary phase factor—depending on time, momentum, or both—may be attached to the Floquet states without affecting $ {{q}^{tt}_{\alpha }}$, thanks to the orthonormality and completeness of the Floquet basis at each instant and each momentum.


\section{Numerical verification of symmetry analysis}
This section provides a numerical study of the symmetry constraints on phase bands and quantum geometry discussed in Sec.~\ref{symsec} of the main text. We introduce the following symmetric model Hamiltonian:
\begin{equation}
\begin{aligned}
&{{H}^{sym}}(\mathbf{k},t)=h^\prime_i~~~\text{for}~~(i-1)T/5<t<(i)T/5,\\
&h_1^\prime=J(\cos (k_x+k_y)+\cos (k_x-k_y)+\cos (-k_x+k_y)\\
&~~~~~~+\cos (-k_x-k_y))\sigma_x,\\
&h_3^\prime=\delta\sigma_z,\\
&h_5^\prime=-J(\cos (k_x+k_y)+\cos (k_x-k_y)+\cos (-k_x+k_y)\\
&~~~~~~+\cos (-k_x-k_y))\sigma_x\\
&h_2^\prime=h_4^\prime=\mathbf{0}.
\end{aligned}
\label{hsym}
\end{equation}
The Hamiltonian (\ref{hsym}) explicitly respects SS (with $u_\mathcal{S}=\sigma_x$), TRS (with $\mathcal{T}=\sigma_z\mathcal{K}$), and PHS (with $\mathcal{C}=\sigma_y\mathcal{K}$). In addition, it is invariant under PS ($u_\mathcal{P}=\mathbf{1}$) and $C_4$   rotations (with the trivial operator  $\mathcal{C}_4=\mathbf{1}$). PPHS, PTS, and reflection symmetry are also fulfilled.

\begin{figure}
\includegraphics[width=\linewidth,trim={0 0 0.25cm 0}, clip]{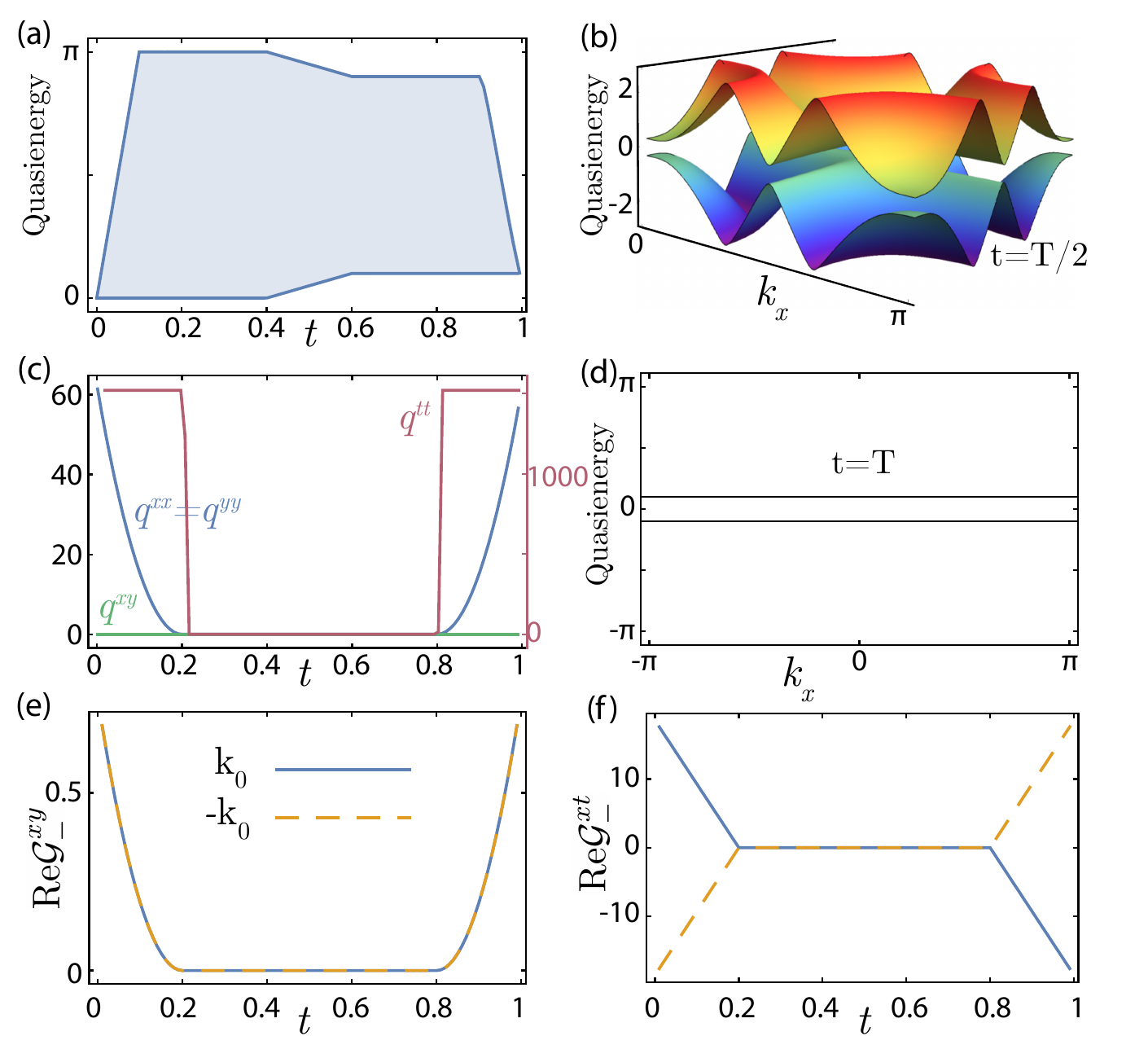} 
\caption{(a) Time dependence of the highest phase band extrema (shaded envelope) for the $H^{sym}$  model. (b) Phase bands at  $t=T/2$ within the first FBZ. (c) Quantum weight components  $q^{mn}$, where $\{m,n\}\in \{x,y,t\}$ with the scale for $q^{tt}$   given on the right axis. The symmetries of the model enforce $q^{xx}=q^{yy}$ and $q^{xy}=q^{xt}=q^{yt}=0$.  (d) Nanoribbon phase bands at  $t=T$. The parameters are $J=3\pi/2, \delta=0, T=1$. (e-f) the nonzero components of quantum geometric tensors $\mathcal{G}_-^{xy}$ and $\mathcal{G}_-^{xt}$ at two special momenta $\mathbf{k}=-\mathbf{k}_0$ and $ \mathbf{k}= \mathbf{k}_0=(0.1,0.2)$. Parameters: $J=5\pi/2, T=1, \delta=\pi/2$.}
\label{hsymfig}
\end{figure}

Fig.~\ref{hsymfig}(a) shows the extrema of the topmost phase band as a function of time for  $J=5\pi/2, \delta=\pi/2, T=1$.  The lower phase band is not displayed because, due to PPHS, it is simply the negative of the topmost band.  The bulk phase bands at $t=T/2$ are shown in Fig.~\ref{hsymfig}(b). For these parameters, the quasienergies are flat with no edge states as implied by Fig.~\ref{hsymfig}(b) where the nanoribbon dispersion is shown.  Both phase bands are symmetric with respect to the momentum and quasienergy axes, owing to PS and PPHS. The quantum metric components are plotted in Fig.~\ref{hsymfig}(c).  As expected from fourfold rotational symmetry and proved in Eq.~(\ref{cnimp}), we observe  $q^{xx}(t)=q^{yy}(t)$ and $q^{xy}(t) =q^{yt}(t)=C^{xt}=C^{yt}=0$.

We now turn to a numerical verification of the symmetry constraints 
on the quantum geometric tensor, which are summarized in 
Table~\ref{symtab}. In particular, according to the PS constraints detailed in Table~\ref{symtab}, we have the following relations:
$\mathcal{G}_{+ }^{xy}(-\mathbf{k},t)=\mathcal{G}_{+ }^{xy}(\mathbf{k},t)$, $\mathcal{G}_{+ }^{xt}(-\mathbf{k},t)=-\mathcal{G}_{+ }^{xt}(\mathbf{k},t)$. Moreover, additional PHS and SS imply  $\mathcal{G}_{+ }^{xy}(\mathbf{k},-t)={\mathcal{G}_{+ }^{xy}}^*(\mathbf{k},t)$, $\mathcal{G}_{+ }^{xt}(\mathbf{k},-t)=-{\mathcal{G}_{ +}^{xt}}^*(\mathbf{k},t)$. This result is fully consistent with Figs.~\ref{hsymfig}(e–f), which concern the behavior of the quantum geometric quantities under the transformation $t\rightarrow -t$. Furthermore, we observe that $\text{Im}\mathcal{G}^{xy}=\text{Im}\mathcal{G}^{xt}=0$ at all times and momenta. This derives from a composite symmetry, which may be termed instantaneous PTS with an anti-unitary operator  $\mathcal{PT}^\prime$ that reverses time, does not change momentum, and satisfies  $\mathcal{PT}^\prime H(\mathbf{k},t) {\mathcal{PT}^\prime}^{-1}=H(\mathbf{k},t)$. 

\section{Proof of some symmetry constraints}\label{appsym}
Here, we provide some constraints imposed by symmetries on phase bands and quantum geometry.

\subsection{sublattice symmetry}
The SS condition can be expressed in terms of the evolution operator as
\begin{equation}
\begin{aligned}
u_\mathcal{S}U(\mathbf{k};0,-t)u_\mathcal{S}^{-1}=U(\mathbf{k},t),
\end{aligned}
\label{}
\end{equation}
where $U(\mathbf{k};0,-t)$ is the evolution operator which evolves the state at time $t=0$ backward in time to $-t$. So, in the presence of SS,
\begin{equation}
\begin{aligned}
&\text{if}~~~U(\mathbf{k},t) \chi_\alpha(\mathbf{k},t)=e^{-i\varphi_\alpha(\mathbf{k},t)} \chi_\alpha(\mathbf{k},t),\\
 &  \text{then}~~~U(\mathbf{k};0,-t){{u}_{\mathcal{S}}}{{\chi }_{\alpha }}(\mathbf{k},t)={{e}^{-i{{\varphi }_{\alpha }}(\mathbf{k},t)}}{{u}_{\mathcal{S}}}{{\chi }_{\alpha }}(\mathbf{k},t) \\ 
&  \Rightarrow U{{(\mathbf{k};-t,0)}^{\dagger }}{{u}_{\mathcal{S}}}{{\chi }_{\alpha }}(\mathbf{k},t)={{e}^{-i{{\varphi }_{\alpha }}(\mathbf{k},t)}}{{u}_{\mathcal{S}}}{{\chi }_{\alpha }}(\mathbf{k},t) \\ 
 &\Rightarrow U(\mathbf{k};-t,0){{u}_{\mathcal{S}}}{{\chi }_{\alpha }}(\mathbf{k},t)={{e}^{i{{\varphi }_{\alpha }}(\mathbf{k},t)}}{{u}_{\mathcal{S}}}{{\chi }_{\alpha }}(\mathbf{k},t) \\ 
& \Rightarrow {{U}_{T}}(\mathbf{k})U{{(\mathbf{k},T-t)}^{\dagger }}{{u}_{\mathcal{S}}}{{\chi }_{\alpha }}(\mathbf{k},t)={{e}^{i{{\varphi }_{\alpha }}(\mathbf{k},t)}}{{u}_{\mathcal{S}}}{{\chi }_{\alpha }}(\mathbf{k},t) \\ 
 & \text{if}\,\,\,\,{{U}_{T}}=1\Rightarrow U(\mathbf{k},T-t){{u}_{\mathcal{S}}}{{\chi }_{\alpha }}(\mathbf{k},t)={{e}^{-i{{\varphi }_{\alpha }}(\mathbf{k},t)}}{{u}_{\mathcal{S}}}{{\chi }_{\alpha }}(\mathbf{k},t), 
\end{aligned}
\label{sslong}
\end{equation}
 confirming the equivalence of phase bands at $t$ and $T-t$ when $U_T=1$.
 
\subsection{particle-hole symmetry}
The PHS condition on the evolution operator takes the form:
\begin{equation}
\begin{aligned}
\mathcal{C}U(-\mathbf{k},t)\mathcal{C}^{-1}=U(\mathbf{k},t).
\end{aligned}
\label{}
\end{equation}
we thus conclude
\begin{equation}
\begin{aligned}
&\text{if}~~~U(\mathbf{k},t) \chi_\alpha(\mathbf{k},t)=e^{-i\varphi_\alpha(\mathbf{k},t)} \chi_\alpha(\mathbf{k},t),\\
&\text{then}~~~U(-\mathbf{k},t) \mathcal{C}\chi_\alpha(\mathbf{k},t)=e^{i\varphi_\alpha(\mathbf{k},t)} \mathcal{C}\chi_\alpha(\mathbf{k},t).
\end{aligned}
\label{}
\end{equation}
This ensures that for every phase band at $\mathbf{k}$, there is another phase band with minus sign at $-\mathbf{k}$.

\subsection{parity symmetry}
The constraint imposed by the PS on the time evolution operator takes the form
\begin{equation}
\begin{aligned}
u_\mathcal{P}U(-\mathbf{k},t)u_\mathcal{P}^{-1}=U(\mathbf{k},t),
\end{aligned}
\label{}
\end{equation}
leading to
\begin{equation}
\begin{aligned}
&\text{if}~~~U(\mathbf{k},t) \chi_\alpha(\mathbf{k},t)=e^{-i\varphi_\alpha(\mathbf{k},t)} \chi_\alpha(\mathbf{k},t),\\
&\text{then}~~~U(-\mathbf{k},t) u_\mathcal{P}\chi_\alpha(\mathbf{k},t)=e^{-i\varphi_\alpha(\mathbf{k},t)} u_\mathcal{P}\chi_\alpha(\mathbf{k},t),
\end{aligned}
\label{}
\end{equation}
showing the symmetry of the phase bands upon inversing the momuntum $\mathbf{k}\leftrightarrow -\mathbf{k}$.

\subsection{$n$-fold rotational symmetry}
 The transformation of the position operator under $\mathcal{C}_3$ rotation is given by $(x^\prime,y^\prime)=\mathcal{C}_3 (x,y) \mathcal{C}_3^{-1}=\frac{1}{2}(-x-\sqrt{3}y, \sqrt{3}x-y)$ and $(x^{\prime\prime},y^{\prime\prime})=\mathcal{C}_3^2 (x,y) {\mathcal{C}_3^2}^{-1}=\frac{1}{2}(-x+\sqrt{3}y, -\sqrt{3}x-y)$. Then we can write 
\begin{equation}
\begin{aligned}
&\mathcal{G}_{+ }^{x y }(\mathbf{k}^\prime)=-\frac{\sqrt{3}}{4} \mathcal{G}_{+ }^{x x} +\frac{\sqrt{3}}{4} \mathcal{G}_{+ }^{yy} +\frac{1}{4} \mathcal{G}_{+ }^{xy}-\frac{{3}}{4} \mathcal{G}_{+ }^{yx},\\
&\mathcal{G}_{+ }^{x y}(\mathbf{k}^{\prime\prime})=\frac{\sqrt{3}}{4} \mathcal{G}_{+ }^{x x} -\frac{\sqrt{3}}{4} \mathcal{G}_{+ }^{yy} +\frac{1}{4} \mathcal{G}_{+ }^{xy} -\frac{{3}}{4} \mathcal{G}_{+ }^{yx},\\
&\mathcal{G}_{+ }^{x x}(\mathbf{k}^\prime)=\frac{1}{4} \mathcal{G}_{+ }^{x x} +\frac{{3}}{4} \mathcal{G}_{+ }^{yy} +\frac{\sqrt{3}}{4} \mathcal{G}_{+ }^{xy} +\frac{\sqrt{3}}{4}\mathcal{G}_{+ }^{yx},\\
&\mathcal{G}_{+ }^{xx}(\mathbf{k}^{\prime\prime})=\frac{1}{4} \mathcal{G}_{+ }^{x x} +\frac{{3}}{4} \mathcal{G}_{+ }^{yy} -\frac{\sqrt{3}}{4} \mathcal{G}_{+ }^{xy} -\frac{\sqrt{3}}{4}\mathcal{G}_{+ }^{yx},\\
&\mathcal{G}_{+ }^{x t}(\mathbf{k}^\prime) =-\frac{1}{2} \mathcal{G}_{+ }^{x t} -\frac{\sqrt{3}}{2} \mathcal{G}_{+ }^{yt},\\
&\mathcal{G}_{+ }^{x  t}(\mathbf{k}^{\prime\prime}) =-\frac{1}{2} \mathcal{G}_{+ }^{x t}+\frac{\sqrt{3}}{2} \mathcal{G}_{+ }^{yt},\\
\end{aligned}
\label{GC3}
\end{equation}
which leads to the constraints in Table~\ref{symtab}.

\section{Proof of charge pumping formula}\label{apppump}
We provide here a proof of charge pumping formula in Eq.~(\ref{pumpx}) of the main text. To do so, let us write the expectation value of the current operator $j_x=\partial_{k_x}H(t)$ for a Floquet state:

\begin{equation}
\begin{aligned}
  & \langle {{j}_{x}}\rangle_\alpha =\iint{\frac{d{{k}_{x}}dt}{2\pi }\langle {{\psi }_{\alpha }}(t)|{{\partial }_{{{k}_{x}}}}H(t)|{{\psi }_{\alpha }}(t)\rangle =} \\ 
 & =\iint{\frac{d{{k}_{x}}dt}{2\pi }\langle {{\phi }_{\alpha }}(t)|{{\partial }_{{{k}_{x}}}}H(t)|{{\phi }_{\alpha }}(t)\rangle =} \\ 
 & =\iint{\frac{d{{k}_{x}}dt}{2\pi }\{{{\partial }_{{{k}_{x}}}}(\langle {{\phi }_{\alpha }}(t)|H(t)|{{\phi }_{\alpha }}(t)\rangle )} \\ 
 & -\langle {{\partial }_{{{k}_{x}}}}{{\phi }_{\alpha }}(t)|H(t)|{{\phi }_{\alpha }}(t)\rangle -\langle {{\phi }_{\alpha }}(t)|H(t)|{{\partial }_{{{k}_{x}}}}{{\phi }_{\alpha }}(t)\rangle  \\ 
 & =\iint{\frac{d{{k}_{x}}dt}{2\pi }\{{{\partial }_{{{k}_{x}}}}(\langle {{\phi }_{\alpha }}(t)|H(t)|{{\phi }_{\alpha }}(t)\rangle )} \\ 
 & -i(\langle {{\partial }_{{{k}_{x}}}}{{\phi }_{\alpha }}(t)|{{\partial }_{t}}{{\phi }_{\alpha }}(t)\rangle -\langle {{\partial }_{t}}{{\phi }_{\alpha }}(t)|{{\partial }_{{{k}_{x}}}}{{\phi }_{\alpha }}(t)\rangle )\}, \\ 
\end{aligned}
\label{proofpump}
\end{equation}
where we use Eqs.~(\ref{schro2}) and (\ref{orthoeq}) of the main text. Note that the first integrand in the last equation of (\ref{proofpump}) vanishes after integration because both the Floquet states and the Hamiltonian are periodic in time and the Brillouin zone.

\bibliographystyle{apsrev4-2}
\bibliography{ref}

\end{document}